\documentclass[aps,prb,tightenlines,twocolumn,superscriptaddress]{revtex4-2}
\usepackage{graphicx,color}
\usepackage{epstopdf}
\usepackage[utf8]{inputenc}
\usepackage{amsmath}
\usepackage{amsfonts}
\usepackage{amssymb}
\usepackage{physics}
\usepackage[caption=false]{subfig}
\usepackage[normalem]{ulem}
\usepackage{bm}
\usepackage[colorlinks=true,allcolors=blue]{hyperref}%
\usepackage{float}

\begin{document}
\title{Absence of Edge States in The Valley Chern Insulator in Moiré Graphene}

\author{Ahmed Khalifa}
\affiliation{Department of Physics \& Astronomy, University of Kentucky, Lexington, KY 40506, USA}
\author{Ganpathy Murthy}
\affiliation{Department of Physics \& Astronomy, University of Kentucky, Lexington, KY 40506, USA}
\author{Ribhu K. Kaul}
\affiliation{Department of Physics \& Astronomy, University of Kentucky, Lexington, KY 40506, USA}

\begin{abstract}
We study the edge spectrum of twisted sheets of single layer and bilayer graphene in cases where the continuum model predicts a valley Chern insulator -- an insulating state in which the occupied moir\'e  mini-bands from each valley have a net Chern number, but both valleys together have no net Chern number, as required by time reversal symmetry. In a simple picture, such a state might be expected to have chiral valley polarized counter-propagating edge states. We present results from exact diagonalization of the tight-binding model of commensurate structures in the ribbon geometry. We find that for both the single-layer and bilayer moir\'e ribbons robust edge modes are generically absent. We attribute this lack of edge modes to the fact that the edge induces valley mixing. Further, even in the bulk, a sharp distinction between the valley Chern insulator and a trivial insulator requires an exact $C_3$ symmetry.
\end{abstract}
\maketitle

\section{Introduction}

Research in moir\'{e} materials has exploded in the past few years since the discovery of superconductivity and Mott insulators in twisted bilayer graphene (TBG)~\cite{Cao2018,Cao20182,doi:10.1126/science.aav1910}. Subsequent experimental studies have shown similar physics in graphene multilayers such as twisted double bilayer graphene (TDBG)~\cite{Shen2020,Liu2020,Cao2020,PhysRevLett.123.197702,He2021} and trilayer graphene on an HBN substrate~\cite{Chen2019,Chen20192,Chen2020}. 

At low energies and small twist angles, in the bulk of a moir\'e system, one makes the continuum approximation \cite{PhysRevLett.99.256802,doi:10.1073/pnas.1108174108}, in which each valley is studied independently and has a moir\'e band structure. The moir\'e  bands in each valley can have  nonzero Chern numbers. However, time-reversal symmetry forces  bands arising in opposite valleys, and related by time-reversal, to have opposite Chern numbers. This raises the interesting possibility that when two such bands in opposite valleys are occupied one  has a valley Chern insulator. Since the Chern number dictates the chirality of edge modes, the valley Chern insulator could, in principle, realize valley-filtered counter-propagating edge modes. However, this picture is questionable, since  it is likely that absent any new mechanism, an edge would produce strong inter-valley scattering. Such inter-valley scattering is expected to cause back scattering between the counter-propagating edge modes resulting in the localization of the would-be edge states. 

Thus, it is somewhat surprising that a recent experiment on twisted double bilayer graphene near charge neutrality~\cite{Wang2022} found non-local transport in this system, which can be  interpreted in terms of robust edge states of the valley Chern insulator phase.  This raises the question of whether there might be some hidden mechanism that suppresses inter-valley scattering at the edge in a moir\'e system. 

To study this question further, we first study the continuum model and find sets of parameters that result in a valley Chern insulator (Sec.~\ref{sec:cont}). Turning to tight-binding models on the lattice, show that in bulk commensurate systems at small angles, a distinction between trivial and valley Chern insulators requires the commensurate structure to have an exact $C_3$ symmetry (which is achieved by twisting around a Carbon site or honeycomb center with AA stacking) in Sec.~\ref{sec3}. Next, we study generic lattice structures in the ribbon geometry and find that protected edge modes are absent in Sec.~\ref{sec4a}. We repeat our calculations for the twisted double bilayer graphene in Sec.~\ref{sec4b} and show that no protected edge modes exist here either. We briefly summarize our results and present our conclusions in Sec.~\ref{sec5}.

\section{\label{sec:cont}Valley Chern insulators in the Continuum Model of TBG}

\begin{figure}[!t]
	\subfloat[\label{CM_zeromass}]{%
		\includegraphics[width=0.23\textwidth]{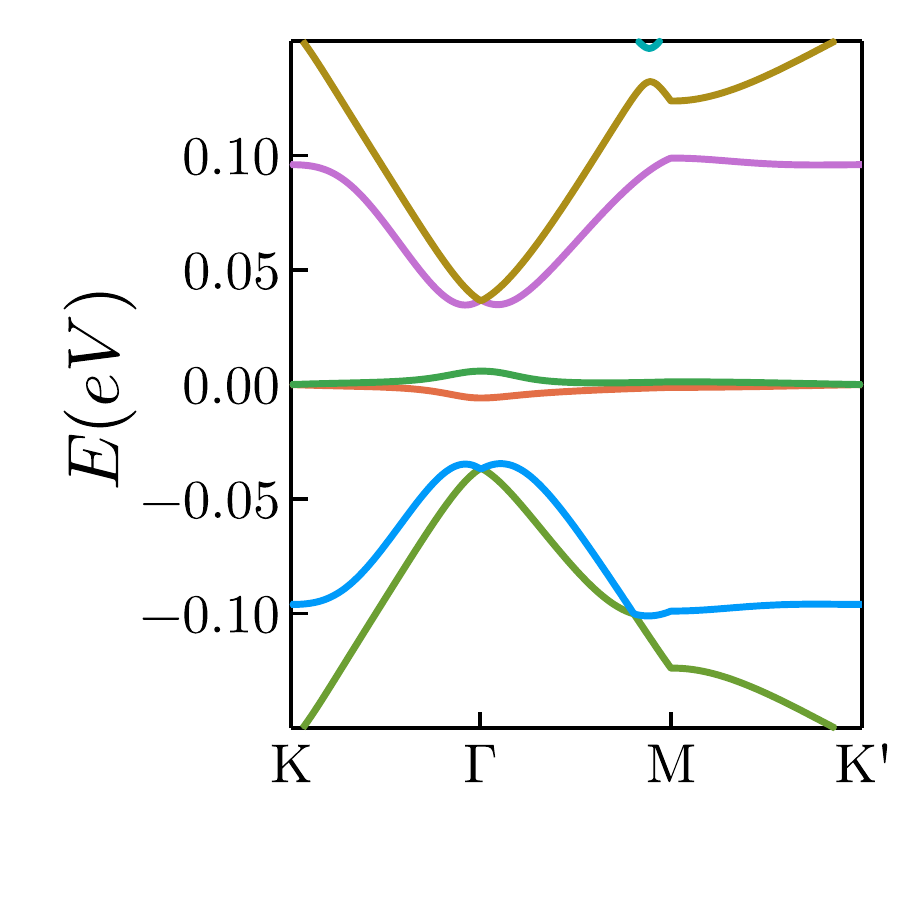}}
	\hspace{0.05em}
	\subfloat[\label{CM-nonzeromass}]{%
		\includegraphics[width=0.23\textwidth]{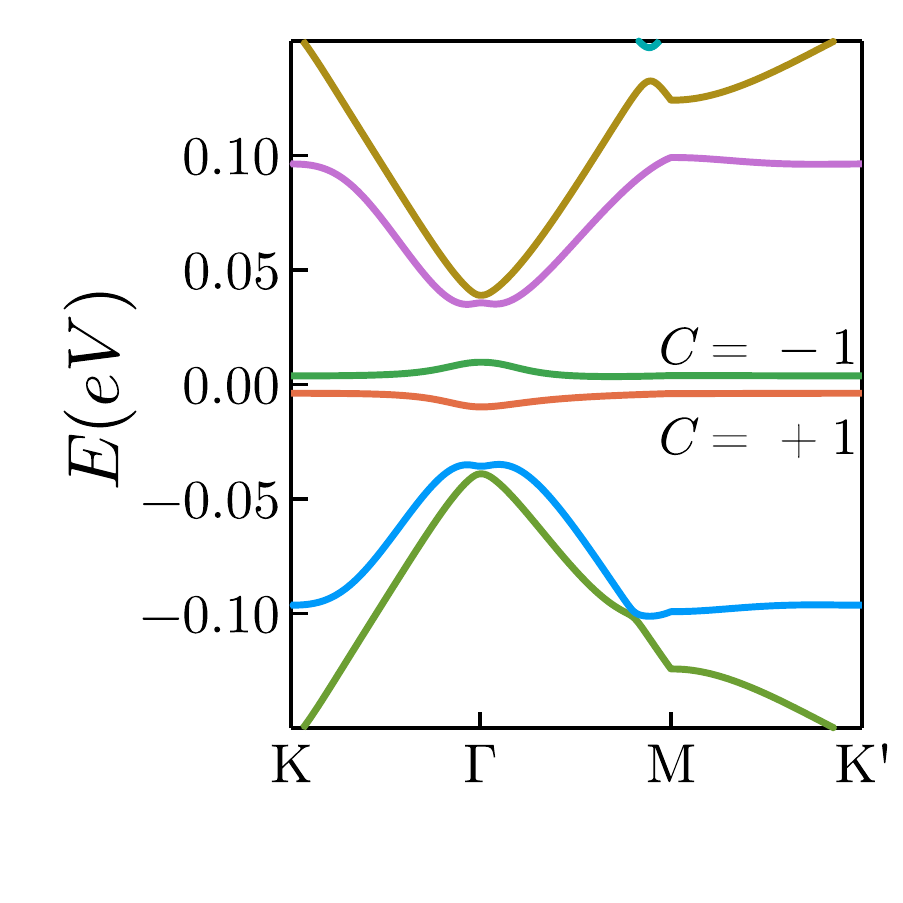}}\\
	\caption{\label{CM_band}Single valley band structure of the continuum model at the magic angle ($\alpha=0.6051$ and $\kappa=0.7$). (a) shows the bands before adding sublattice masses to the layers and the Dirac touchings at the  $K, K'$ points protected by the $C_{2}\mathcal{T}$ symmetry. (b) A gap opens up at the  $K, K'$ points when sublattice masses are added (We used $m_t=0.01, m_b=0.01 eV$). The conduction and the valence bands acquire opposite Chern numbers. By time-reversal symmetry, bands in the other valley have the opposite Chern numbers as their time-reversed partners.}
\end{figure}

A good starting point for studying  moir\'e  physics is the continuum model of twisted bilayer graphene  \cite{PhysRevLett.99.256802,doi:10.1073/pnas.1108174108}. This model, which is well-justified for small twist angles for bands near the charge neutral point, is based on the tunneling between the two Dirac cones (in a single valley) of the two graphene layers. In real space, the single valley continuum model reads,
\begin{equation}\label{BM_model}
H_{\rm cont}(\mathbf{r}) = \begin{pmatrix}
-iv_{F}\bm{\sigma}\cdot\bm{\nabla}_{\theta/2} & U(\mathbf{r})\\
U^{\dagger}(\mathbf{r}) & -iv_{F}\bm{\sigma}\cdot\bm{\nabla}_{-\theta/2}
\end{pmatrix},
\end{equation}
where $\bm{\sigma}=(\sigma_x,\sigma_y)$ is a Pauli matrix vector representing the sublattice index of graphene. The diagonal elements of the matrix represent the low energy continuum Hamiltonian near the Dirac point within each layer. Note that the gradient term is rotated in each layer by $\pm\theta/2$. The graphene Fermi velocity $v_F$ is expressed in terms of the nearest neighbor hopping strength $t$, as $v_F=\frac{-\sqrt{3}ta}{2\hbar}$, where $\hbar$ is set to 1 and $a$ is the graphene lattice constant which we also set to 1 so that $v_F$ has the units of energy. For the following, we use the numerical value $t=-2.7eV$ to obtain the band structure (Fig. \ref{CM_band}). The tunneling between the layers is encapsulated in the $U(\mathbf{r})$ term which reads, \begin{equation}\label{U(r)}
    U(\mathbf{r}) = \sum_{j=1}^{3} T_{j}e^{i\mathbf{q}_{j}.\mathbf{r}},
\end{equation} 
where $T_{j} = w_0+w_{1}e^{(i2\pi/3)j\sigma_z}\sigma_{x}e^{(-i2\pi/3)j\sigma_z}$ with $w_0,w_1$ denoting the strength of AA and AB hopping, respectively and $q_1=(0,-k_\theta), q_2=C_3q_1, q_3=C_3q_2$ where $k_\theta$ is the difference between the rotated Dirac points in each layer. Lattice relaxation effects generally make the AB hopping more favorable than AA hopping~\cite{lattice_relaxation,PhysRevB.99.195419}. This is incorporated into the model by making $w_0 < w_1$ with $\kappa=w_0/w_1$ denoting the ratio between the two. Taking the extreme limit $w_0 = 0$ leads to the chiral model of TBG~\cite{PhysRevLett.122.106405}. The rotation of the Dirac Hamiltonian in each layer is usually ignored at small twist angles due to the fact that $\theta$ is small($\approx0.01$). When this approximation is made, the continuum model becomes a function of a dimensionless parameter $\alpha=\frac{w_1}{v_{F}k_\theta}$ and the ratio $\kappa$. The most studied feature of the continuum model is a pair of nearly flat (exactly flat in the chiral model) bands (per valley per spin) at the magic angles~\cite{PhysRevB.82.121407,doi:10.1073/pnas.1108174108}. These bands are isolated from high energy bands by an energy gap.

The single-valley continuum model has several emergent symmetries that constrain its band structure and its topological properties~\cite{PhysRevX.8.031089,PhysRevB.98.085435,PhysRevX.8.031088}. Of particular interest is the $C_{2}\mathcal{T}$ symmetry which protects a Dirac touching at the $K, K'$ points of the moir\'e  BZ (see Fig. \ref{CM_zeromass}). Upon breaking this symmetry a gap opens up at the $K, K'$ points. This symmetry-breaking can be achieved experimentally by aligning the graphene layers with the hexagonal Boron Nitride (HBN) substrate as has been shown before in the single layer\cite{Jung2015,PhysRevB.96.085442,doi:10.1126/science.1237240,PhysRevLett.110.216601,Zibrov2018,Kim2018,PhysRevB.92.155409}. The effect of the HBN substrates is modelled by adding a uniform $C_2$ breaking mass on the top and the bottom layer. Thus, equation (\ref{BM_model}) becomes; \begin{equation}\label{BM_model_mass}
    H(\mathbf{r}) = 
H_{\rm cont}(\mathbf{r})+
\begin{pmatrix}
m_{t}\sigma_z & 0\\
0 & m_{b}\sigma_z
\end{pmatrix}.
\end{equation}   
 In addition to opening up a gap at the $K, K'$ points, the two flat bands can acquire  nonzero Chern numbers. Time-reversal forces the Chern numbers of corresponding bands in opposite valleys to  have opposite Chern numbers (see Fig. \ref{CM-nonzeromass}). Adding the effects of electron-electron interactions to this one-body picture can explain the observation of  the quantum anomalous hall effect in TBG devices that are aligned with HBN\cite{doi:10.1126/science.aay5533,doi:10.1126/science.aaw3780,Lu2019}. In brief, the quantum anomalous Hall effect can be understood as the result of spontaneously breaking time-reversal symmetry when the electrons fill a single-valley flat band with nonzero Chern number\cite{PhysRevLett.124.166601,PhysRevB.99.075127,PhysRevResearch.1.033126,PhysRevX.10.031034,PhysRevX.9.031021}.\\ 

As described before, the valley Chern insulator is a time reversal symmetric band insulator in which each valley has a non-zero Chern number (but the two valleys have opposite Chern numbers by time reversal). The valley Chern number is defined by integrating the Berry curvature of one valley on the moir\'e Brillouin zone. In the continuum model, the valley Chern number in the TBG/HBN system is  well-defined  due to the conservation of charge in each valley. We note that unlike the quantum anomalous Hall phase, this state does not require electron-electron interactions. The continuum model of the TBG/HBN structure suggests that the system can realize a valley Chern insulator phase at charge neutrality in which the two valleys have opposite valley Chern numbers.\\

\begin{figure}[!t]
	\subfloat[\label{alpha0.2}]{%
		\includegraphics[width=0.42\linewidth]{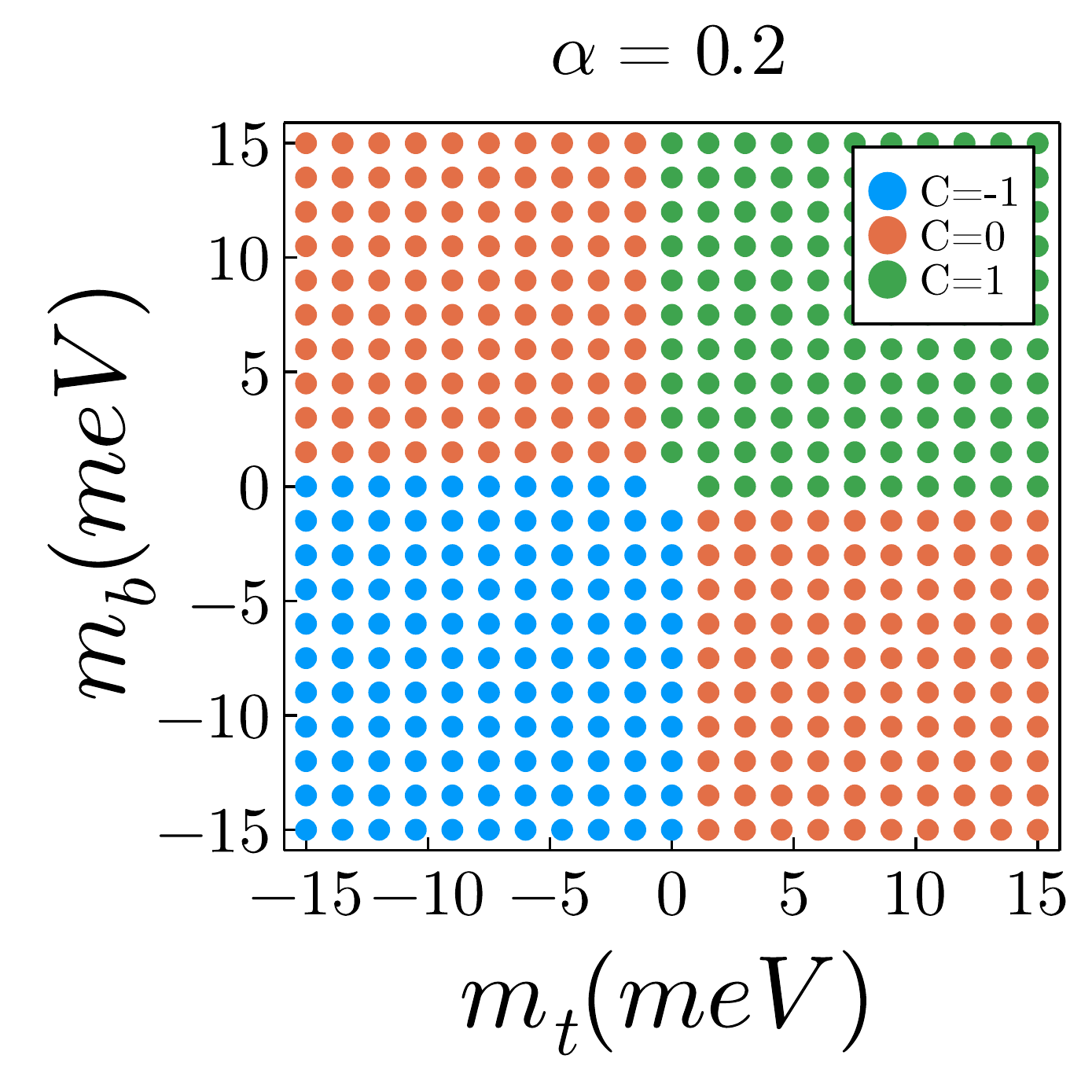}}
	\hspace{0.05em}
	\subfloat[\label{alpha0.4}]{%
		\includegraphics[width=0.42\linewidth]{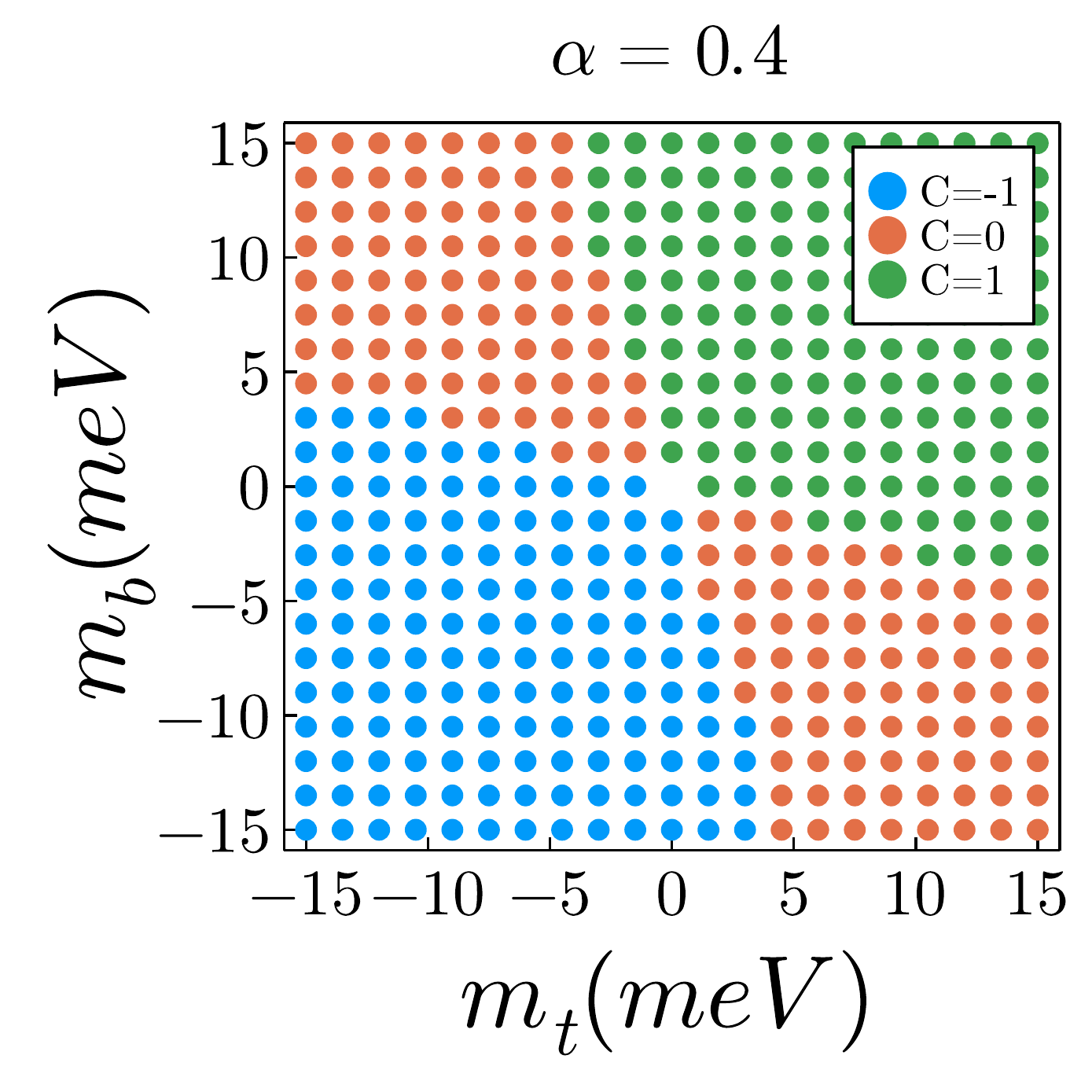}}
		\hspace{0.05em}
	\subfloat[\label{alpha0.6051}]{%
		\includegraphics[width=0.42\linewidth]{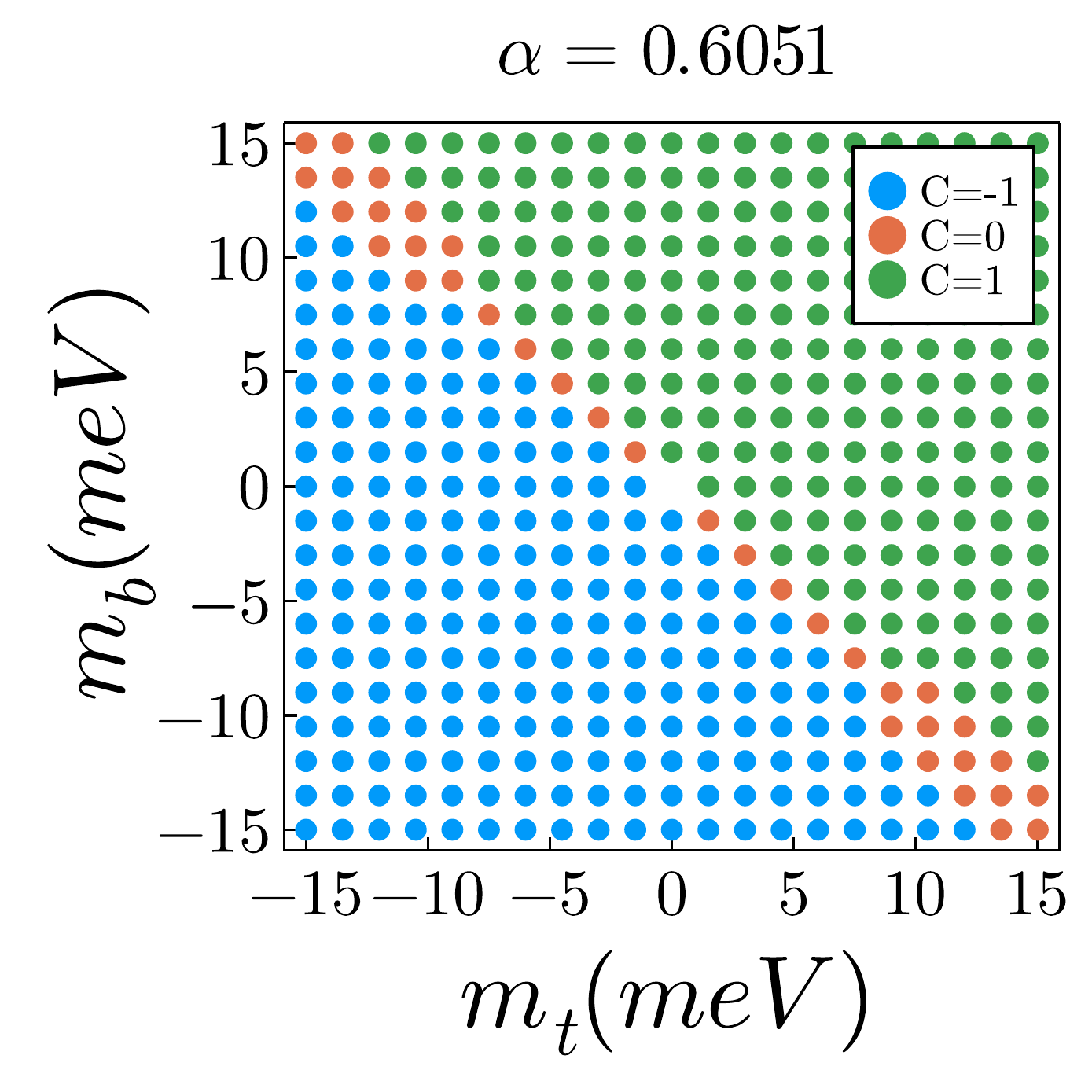}}
		\hspace{0.05em}
		\subfloat[\label{alpha0.65}]{%
		\includegraphics[width=0.42\linewidth]{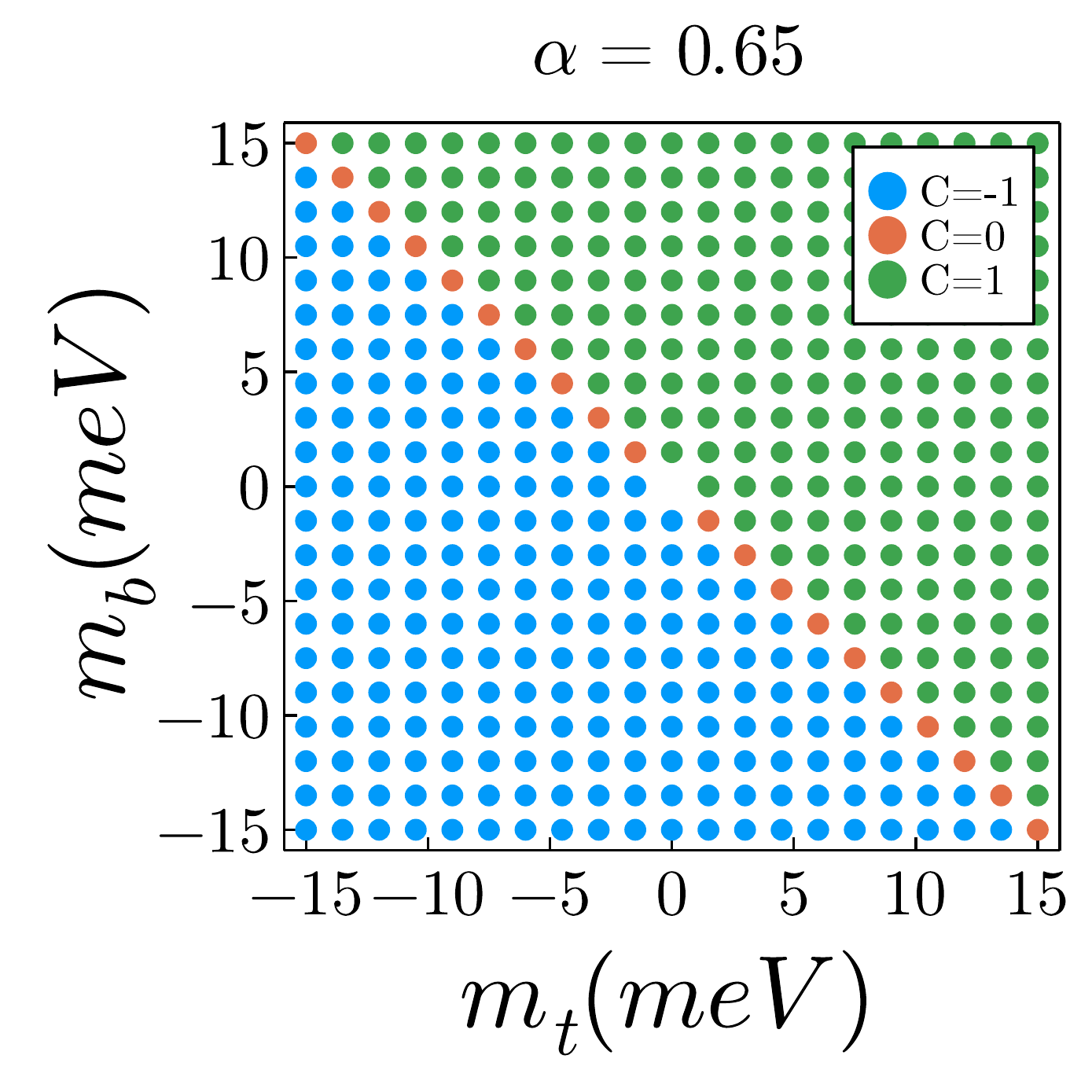}}\\
	\caption{\label{CM_phasediagaram}Phase diagram of the valley Chern insulator phase in the continuum model. The Chern number is calculated for the conduction band near charge neutrality. At small interlayer coupling (small $\alpha$), the quadrants with the sublattice masses on top and bottom layer ($m_t,m_b$) having opposite signs are in a trivial phase with zero Chern number (a), (b). As the interlayer coupling increases, the strong renormalization of the bands leads to shrinkage of the trivial phase and to an enhancement of the valley Chern insulator. we obtained these figures using $\kappa=0.7$ and we get a qualitatively similar phase diagram in the chiral limit ($\kappa=0$).}
\end{figure}

We map out the valley Chern number phase diagram in the continuum model for different values of $\alpha$. We see that at small $\alpha$, which corresponds to weak interlayer hopping, the Chern number is zero whenever the two masses have opposite signs.  This is to be expected due to the cancellation of the Berry phase between the two layers. As we approach the strong coupling regime (large $\alpha$) the  zero Chern number region shrinks in the phase diagram, and the system is mostly in the valley Chern insulator phase (see Fig. \ref{CM_phasediagaram}).
\section{\label{sec3}Studying Commensurate lattices in real space}
\subsection{Geometry}

\begin{figure}
    \includegraphics[width=0.9\linewidth]{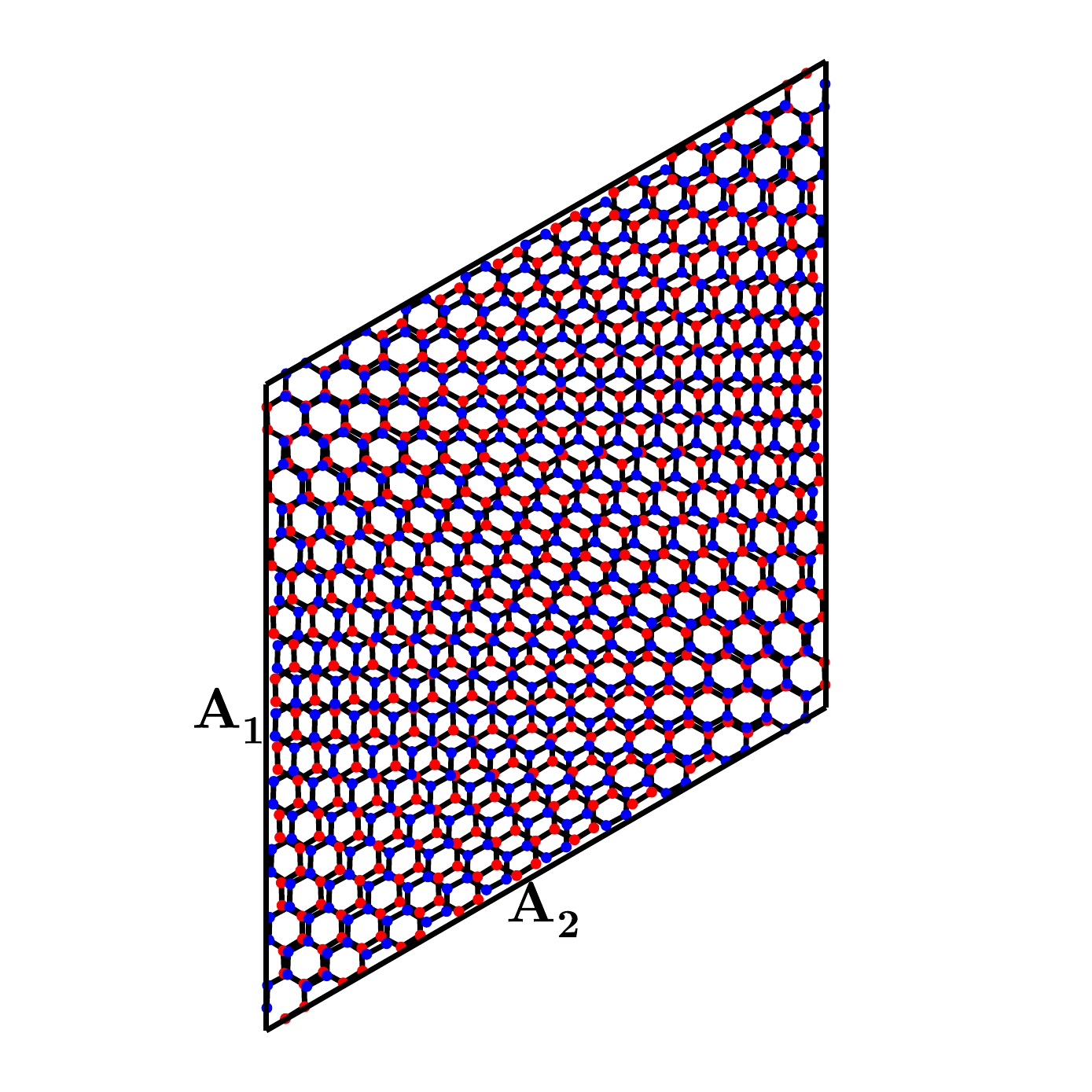}
    \caption{An example moir\'e unit cell of TBG with a commensurate angle $\theta\approx 3.48^\circ (m=10,n=9)$ with 1084 atoms in the unit cell. $\mathbf{A_1}$ and $\mathbf{A_2}$ are the moir\'e  lattice translation vectors. A ribbon with a generic edge can be constructed by taking $\mathbf{A_2}$ to be infinite and a number of moir\'e  unit cells in the $\mathbf{A_1}$ direction defines the width of the ribbon.}
    \label{moire_unitcell}
\end{figure}

In a lattice model or a real sample of graphene, the individual valley charges are not conserved. Only the total charge is conserved and hence states in different valleys can mix. In order to study the effect of valley mixing we study TBG via a tight binding model in real space. 
 This gives us two handles on the problem. First we are able to study the mixing between the valleys, and secondly we have the means to investigate the edge states that are associated with the boundaries of any putative topological phases. We first review the geometry of commensurate TBG which is constructed from two layers of honeycomb lattices that are aligned on top of each other (AA stacking) then rotated with a relative angle $\theta$ and a possible translation between the layers. The system is not periodic for a general twist angle $\theta$.  However, for a set of discrete twist angles, the system has exact translation symmetry with an enlarged super cell\cite{PhysRevLett.99.256802,PhysRevB.81.161405,PhysRevB.81.165105} (see Fig. \ref{moire_unitcell}). This set of discrete twist angles is defined by co-prime positive integers $(m,n)$ that is given by the formula: 
\begin{equation}\label{commensurate_condition}
    \cos{\theta}=\frac{m^2+n^2+4mn}{2(m^2+n^2+mn)}
\end{equation}
We assume that the top sheet rotates by angle $\theta/2$ and the bottom sheet by $-\theta/2$, the translation vectors for the superlattice are given by~\cite{PhysRevB.87.205404} 
\begin{equation}\label{superlattice_vecs}
\mathbf{A}_1=n\mathbf{a}^t_1+m\mathbf{a}^t_2, \mathbf{A}_2=-m\mathbf{a}^t_1+(m+n)\mathbf{a}^t_2,
\end{equation}
where $\mathbf{a}^t_{1,2}=R(\theta/2)\mathbf{a}_{1,2}$ are the translation vectors of the top graphene layer with $R(\theta/2)$ being the 2D matrix of rotation and $\mathbf{a}_{1,2}$ are the unrotated graphene lattice translation vectors. The length of the supercell lattice translation vectors is expressed in terms of the rotation angle as,
\begin{equation}\label{length_vecs}
    A\equiv\abs{\mathbf{A}_1}=\abs{\mathbf{A}_2}=\frac{(m-n)a}{2\sin{\theta/2}},
\end{equation}
where $a$ is the length of the graphene lattice constant and it is assumed that $m>n$. We note that when $m-n=1$ the length of the commensurate superlattice vector coincides with that of the emergent moir\'e  pattern in the continuum. One can see that the moir\'e pattern grows inversely with the angle $\theta$ and for a given $m,n$ the number of atoms in the unit cell ($N$) is given by $N=4(m^2+mn+n^2)$. 

We conclude this subsection by considering the rotational symmetry of the lattice. We start with the two layers in perfect registry. In general, the center of rotation can be any point in space. However, we restrict ourselves to either rotating about a Carbon site or about the center of a hexagon. Rotating about these special points leads to the system either having $C_6$ symmetry when rotating about the center of a hexagon~\cite{PhysRevB.81.161405,PhysRevLett.123.036401,PhysRevB.98.085435} or a $C_3$ symmetry when the center of rotation is a Carbon site. These symmetries are removed upon translating one of the layers in a generic direction after applying the twist.   
\subsection{Tight-Binding Model}
We now study TBG in real space via a tight-binding model based on the the Carbon $p_z$ orbitals tight binding models first introduced in graphene and then further generalized to graphene hetero structures in previous studies\cite{T2010,PhysRev.94.1498,PhysRevB.87.205404,PhysRevB.87.075433}. The Hamiltonian is given by,  
\begin{equation}\label{TB_Ham}
    H=\sum_{\mu=1,2}\sum_{i,j}t_{ij}c_{\mu,i}^{\dagger}c_{\mu,j}+\sum_{i,j}t_{ij}^{\perp}c_{1,i}^{\dagger}c_{2,j}+\text{h.c.},
\end{equation}
where $c_{\mu,i}$ is annihilation operator for an electron in layer $\mu$ and at lattice position $i$. The first term accounts for the hopping within each graphene layer where we use the numerical value of $t=-2.7eV$ for the nearest neighbor intralayer hopping strength, and set the further neighbor intralayer hoppings to zero. The second term accounts for the interlayer hopping. In order to make a straightforward correspondence between the continuum and  lattice models, we assume a simple form for the interlayer hopping that is a function of the in-plane distance between the two atoms (in different layers) and decays exponentially with increasing distance. The form assumed for the interlayer hopping $t_{ij}^{\perp}$ is $t^{\perp}(r)=t_{v}\exp(-r/\xi)$, where $r$ is the in-plane distance between the two atoms with $\xi$ being the decay constant and $t_v$  the hopping amplitude. To match the lattice calculations with the continuum model, we recall the continuum model is a function of $\alpha=\frac{w_1}{v_{F}k_\theta}$, and for the form of interlayer hopping chosen, $w_1$ is given by, $w_1=\frac{2\pi\xi^2}{\Omega_{u.c}(1+k_{D}^2\xi^2)^{3/2}}t_v$ (See Appendix \ref{app:lattcont} for the derivation).
For a given $\alpha$ and $\xi$ we solve for the value of $t_v$ that enters in the tight binding model. The lattice relaxation effects are incorporated in the same way as in the continuum model, by reducing the $AA$ hopping strength compared to the $AB$ hopping. Finally, the HBN substrate is modeled by adding a sublattice mass term on each layer to our tight binding Hamiltonian(equation (\ref{TB_Ham})). A comparison between the band structures of the lattice model and the continuum theory has been carried out previously \cite{PhysRevB.87.205404,PhysRevB.99.205134,PhysRevB.101.195425},  but is provided in Appendix \ref{app:lattcont} for completeness. We thus have a controlled correspondence between the continuum model and the commensurately twisted lattice model. For this work, a central feature  of the band structure in the commensurate super cell structure is that there is no separation between the valleys as there was in the continuum model. Indeed the bands from both valleys of the continuum model can mix and appear superposed in the moir\'e Brillouin zone. This renders the simple definition of the valley Chern number in the continuum model ill-defined in the commensurately twisted lattice model. 

\begin{figure}[!t]
	\subfloat[\label{gap}]{%
		\includegraphics[width=0.23\textwidth]{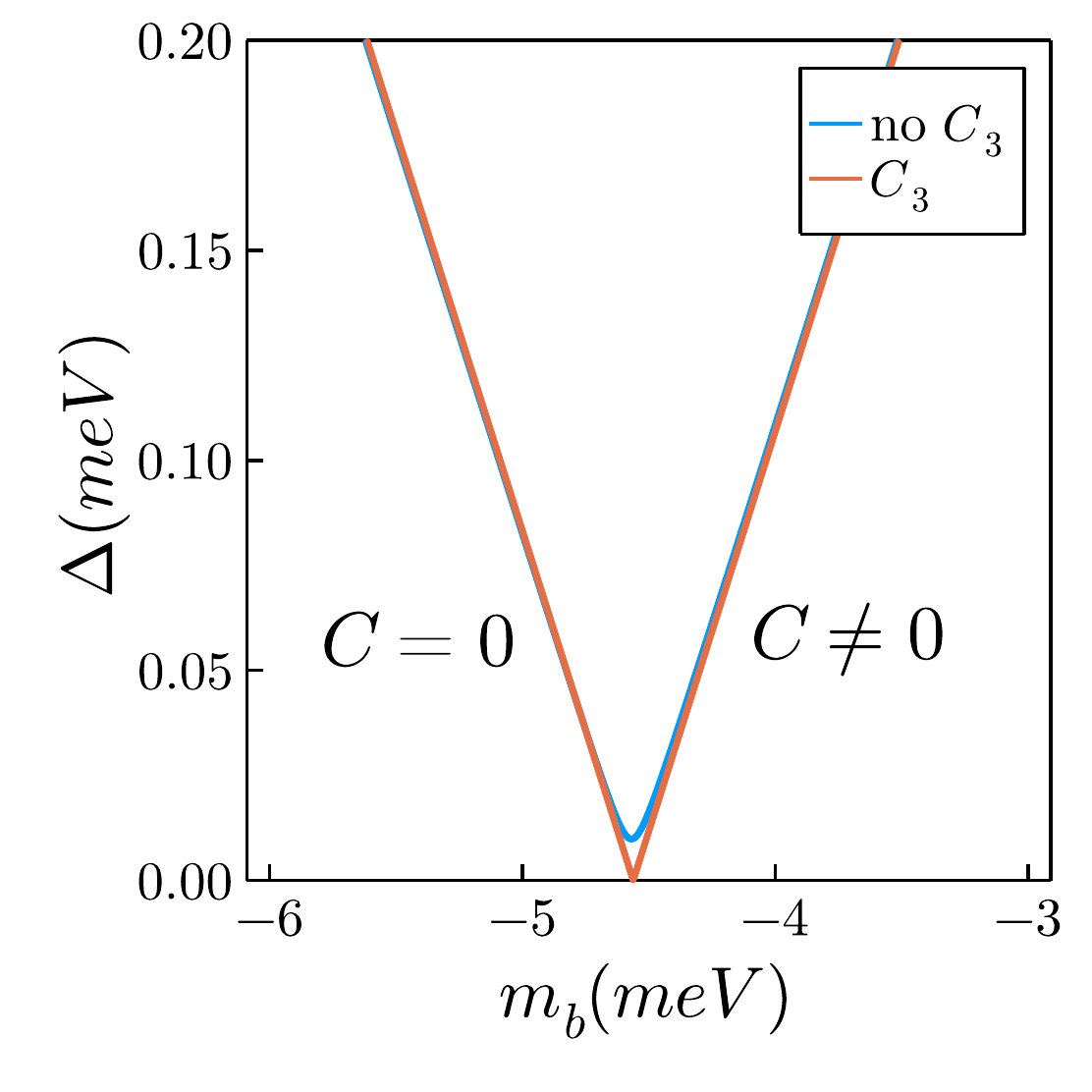}}
	\hspace{0.05em}
	\subfloat[\label{C3eig}]{%
		\includegraphics[width=0.23\textwidth]{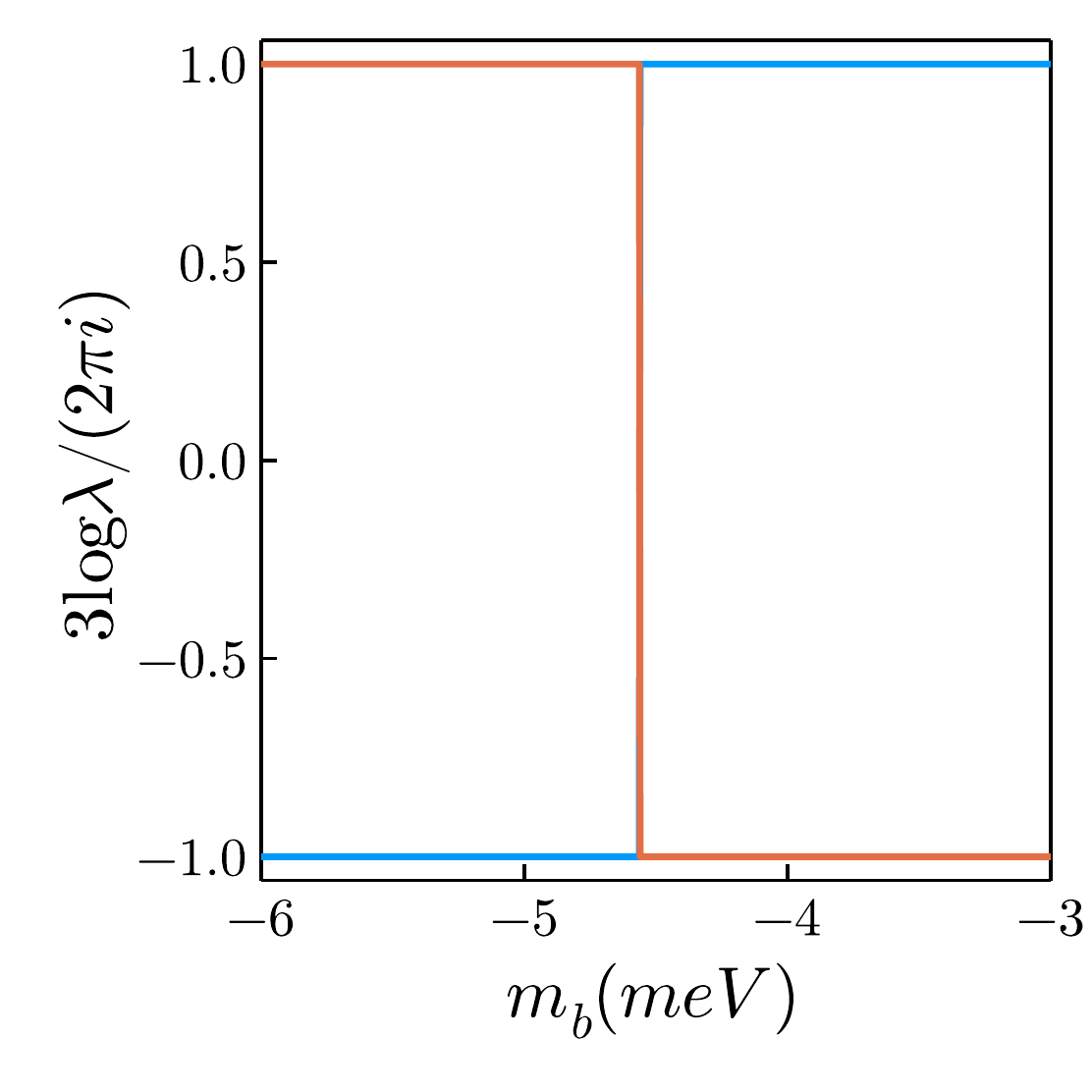}}\\
	\caption{\label{topological_transition}Spectrum at the $K$ point obtained by the lattice tight binding model. (a) shows the evolution of the gap between the middle bands as we vary the Hamiltonian parameters ($m_t,m_b$) going through the topological transition defined by the continuum model. The path is set by $m_t=10meV$ and varying $m_b$ from $-6meV$ to $-3meV$. A gap only closes when $C_3$ symmetry is present on the lattice. When an exact $C_3$ symmetry is absent the gap does not close and hence the trivial and valley Chern insulator phases of the continuum are smoothly connected on the lattice.(b) When $C_3$ symmetry is present, the two bands touching at the $K$ point exchange their $C_3$ eigenvalues across the topological transition. The two figures were obtained using $\theta\approx1.08$($m=31,n=30$), $\alpha=0.6051$, $\kappa=0.7$ and $\xi=0.3a$. The graphene lattice constant ($a$) is set to 1.}
\end{figure}

In the continuum model there is a sharp topological transition between the  trivial and valley Chern insulator phases, that takes place simultaneously but independently in each valley (in the presence of time reversal symmetry). Since this transition is accompanied by a change in the Chern number in the continuum model, a gap closing must occur in the spectrum (specifically  at the $K$ and $K'$ points~\cite{PhysRevLett.124.166601}). We now consider how the band structure evolves in the commensurate system as the corresponding continuum approximation is taken from a phase with zero valley Chern number to a one with nonzero valley Chern number (see Fig.~\ref{CM_phasediagaram}).  The fact that the gap closing happens at the $K$ and $K'$ points suggest that this is due to the emergent $C_3$ symmetry in the continuum model. In the lattice, however, $C_3$ symmetry is only present when the center of rotation is a center of a hexagon or a Carbon site but is removed upon carrying out a subsequent generic translation. The spectrum at the $K$ point indeed shows a gap closing at the transition point only when $C_3$ symmetry is present (see Fig. \ref{gap}). Furthermore, when $C_3$ symmetry is present, bands can be labeled by the $C_3$ eigenvalues (which we denote by $\lambda$) where $\lambda$ can take any value of the cubic roots of unity ($1,\omega,\omega^2$). The two bands that touch at the transition point exchange their $C_3$ eigenvalues (see Fig. \ref{C3eig}). In the absence of $C_3$ symmetry (achieved in our simulation by rotating and then carrying out a generic translation) we find that the gap does not close and the two phases are connected to each other smoothly (see Fig. \ref{gap}).

\section{\label{sec4}Edge states in a ribbon geometry}
\subsection{\label{sec4a}TBG}

We now turn to the motivating question of our study, whether the moir\'e system when in a regime that has a valley Chern insulator in the continuum, has edge states when studied in a lattice system with a boundary.
Our tight-binding model allows us to get the band structure in a ribbon geometry, thus allowing access to the edge states of the TBG aligned with HBN system. An infinite ribbon can be constructed by taking the system to be infinite along one of the lattice primitive lattice vectors ($\mathbf{A}_2$) (with momentum being a good quantum number along this direction) and finite in the other direction with the width of the ribbon set by the number of unit cells along the finite direction. Since the number of atoms in the unit cell in the  ribbon geometry at small angles is quite large (the number of atoms in one moir\'e unit cell is $\sim10,000$ atoms), we only obtain the bands near charge neutrality where the most interesting physics occurs.

Upon diagonalizing this Hamiltonian, we obtain two kinds of states; those that are inherited from the infinite system (bulk bands) and states that are localized at the edge (edge states). We determine the character of the states by computing the position expectation value of the wave function along the width of the ribbon\cite{PhysRevB.103.155410}. We then give each state a color that labels its character, with green denoting bulk states, and red/blue denoting states localized on the top and bottom edge, respectively.

\begin{figure}[!t]
	\subfloat[\label{nonchiral}]{%
		\includegraphics[width=0.49\linewidth]{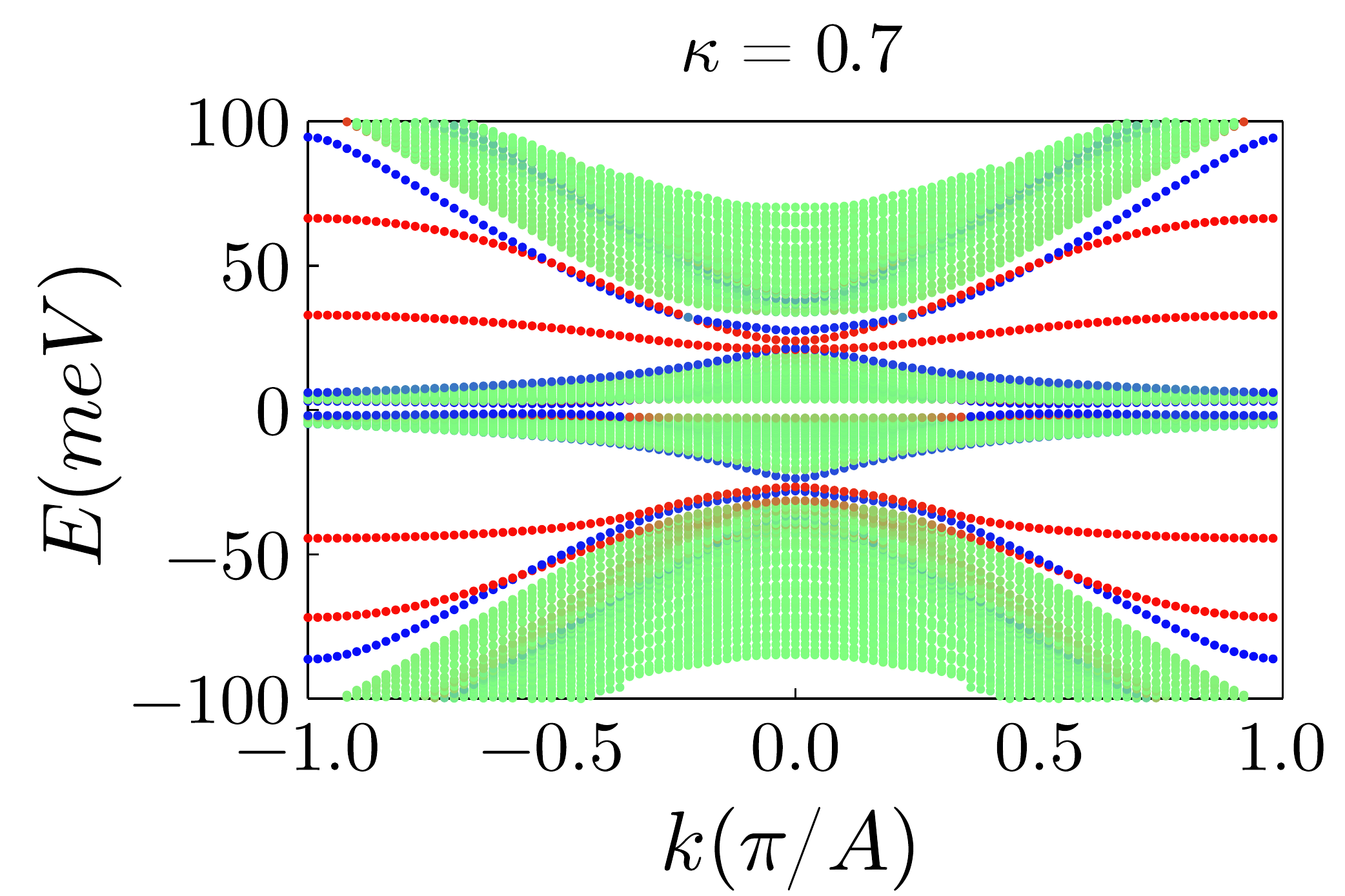}}
	\hspace{0.05em}
	\subfloat[\label{chiral}]{%
		\includegraphics[width=0.49\linewidth]{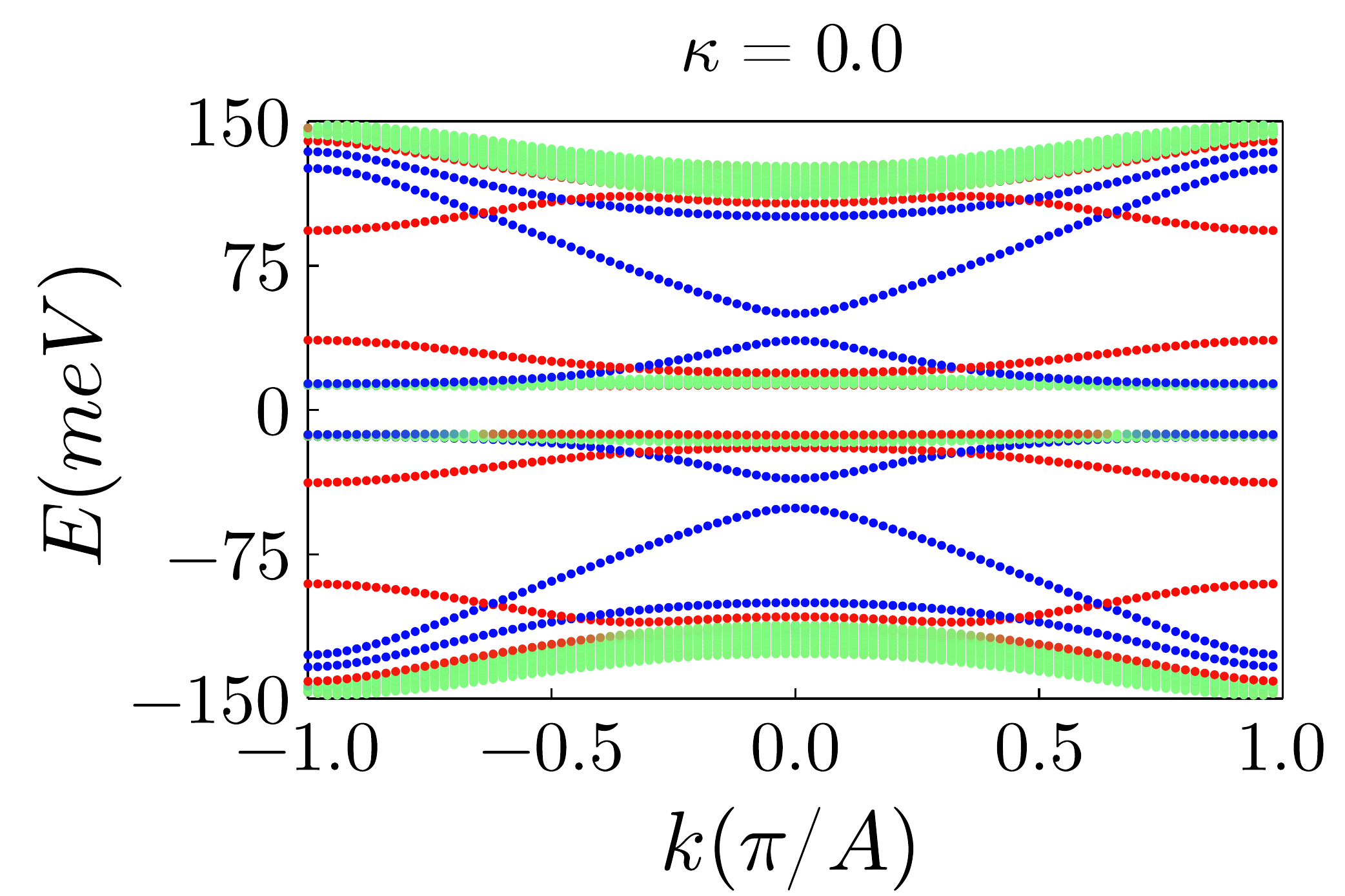}}
		\hspace{0.05em}
	\subfloat[\label{nonchiral_zoom}]{%
		\includegraphics[width=0.49\linewidth]{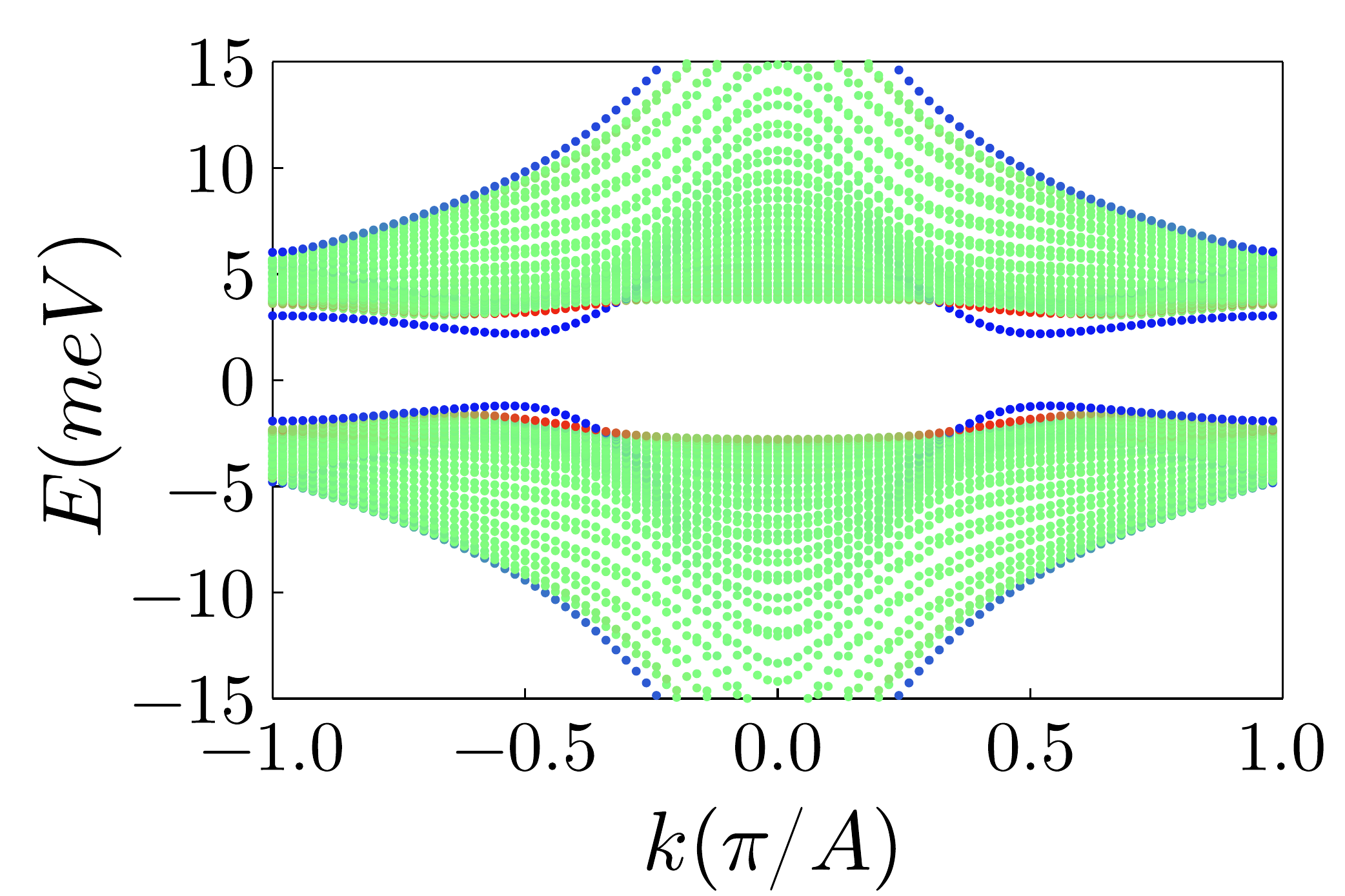}}
		\hspace{0.05em}
		\subfloat[\label{chiral_zoom}]{%
		\includegraphics[width=0.49\linewidth]{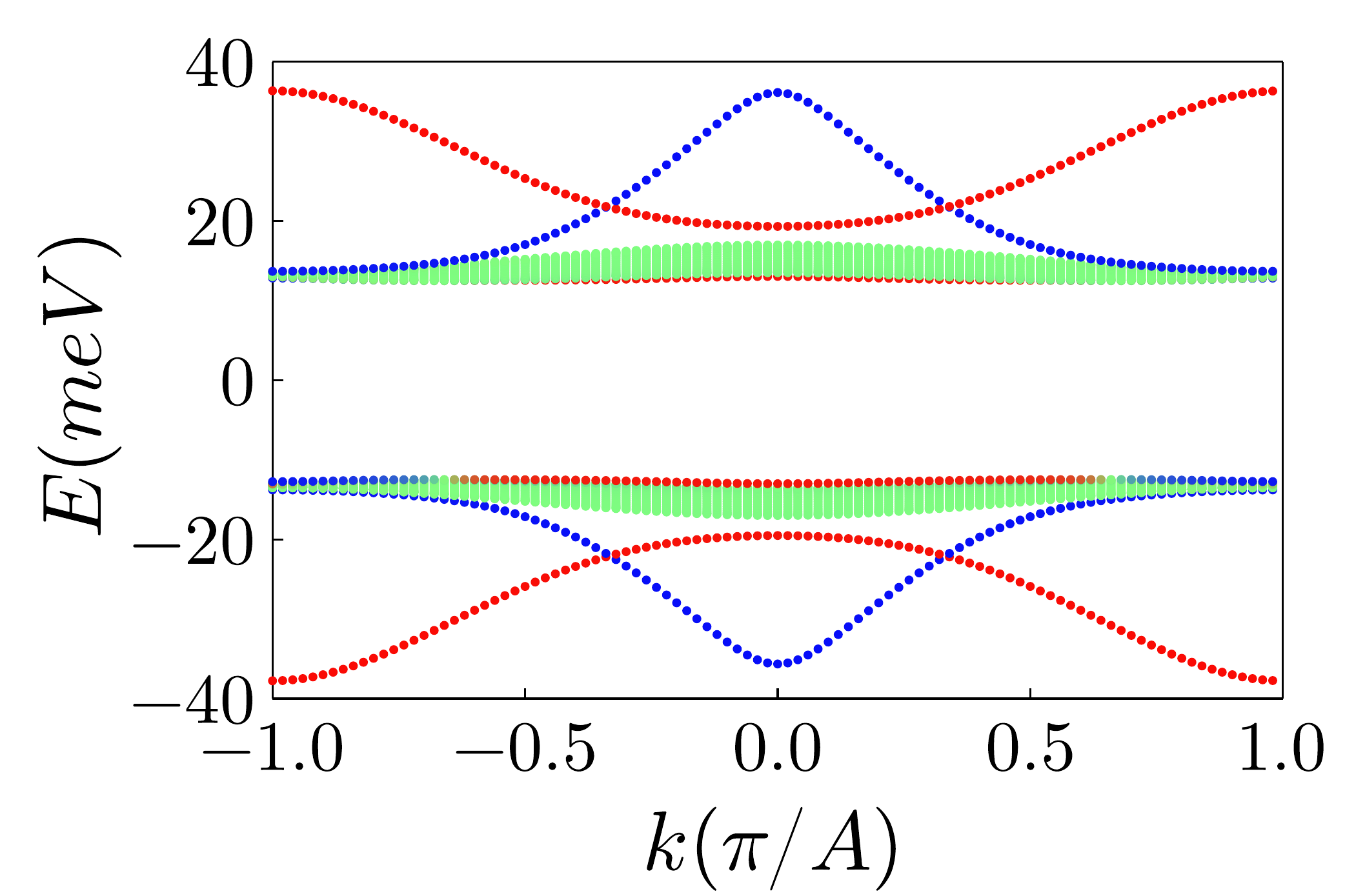}}\\
	\caption{\label{TBG_ribbon}Band structure of a TBG ribbon with a width of 20 moir\'e unit cells. We used $m=31$ and $n=30$ which corresponds to a commensurate angle $\theta\approx1.08^\circ$ and further a generic translation between the layers is applied. The value of $\xi=0.3$ is used. The color of the energy levels are assigned based on the average transverse location of the states-- the green states are delocalized and the red and blue states are localized on the top and bottom edge respectively. (b),(d) were obtained at the chiral limit where (a) and (c) were obtained with $\kappa=0.7$ The lower panels are detailed views of the upper panels in the region of energies corresponding to charge neutrality. Both layers have sublattice masses of magnitude $m_t=15$ meV and $m_b=10$ meV which puts the system in the valley Chern insulator phase in the continuum model. Absence of dispersing gap crossing edge modes at charge neutrality is shown as we zoom in on the bands (c),(d).}
\end{figure}

We now discuss our findings for the ribbon's band structure in the chiral and the non-chiral limits. Our results for the ribbon band structure are shown in Fig.~\ref{TBG_ribbon}. We have studied the spectrum both for $\kappa=0$ (b,d), the chiral limit and the more generic value of $\kappa=0.7$ (a,c). The lower panels are zoom-ins of the upper panels to emphasize the behavior close to charge neutrality. First, we find edges states, blue and red states that occur between the flat bulk bands and the higher energy bulk bands. We identify these as the ``moir\'{e} edge states" which were previously studied in the literature~\cite{PhysRevB.87.075433,PhysRevB.89.205405,PhysRevB.91.035441,PhysRevB.99.155415,PhysRevB.97.205128,PhysRevB.103.155410}. Since these do not occur between the flat bands they will not affect the low-energy behavior at the edge at charge neutrality, which is the focus of this study. Most strikingly for our study, even when the system has mass parameters that lie in the valley Chern insulator phase within the continuum model, we find a clear absence of dispersing edge states that connect the two flat bands that acquire a nonzero Chern number in the continuum model. We note that we have obtained the same result with different values of $m_t$ and $m_b$ and for different types of edges. We recall that in a naive interpretation of the continuum model, a valley Chern insulator should have valley polarised edge states with each valley contributing state of opposite chirality. Our finding appears in contradiction to this expectation. We identify two reasons for this absence. First, inter-valley scattering is generically present at the edge, and secondly, $C_3$ symmetry (whose importance was established in the previous section) is generally broken at the edge. Either effect is enough to gap out the would-be edge states associated with the valley Chern insulator in TBG.
\subsection{\label{sec4b}TDBG}
In this subsection we carry out a similar tight binding calculation for the TDBG system which was the subject of the experimental study in \cite{Wang2022}. A  continuum model \cite{PhysRevLett.99.256802,doi:10.1073/pnas.1108174108,PhysRevB.99.235417,PhysRevB.100.201402,PhysRevB.99.235406} can be formulated to study the bands of TDBG in the same way as for TBG. The single valley TDBG Hamiltonian obtained by twisting two bilayers of AB stacked graphene is given by, 
\begin{equation}
H(\mathbf{r})=\begin{pmatrix}
H^{t}_{AB} & \Tilde{U}(\mathbf{r})\\
\Tilde{U}^{\dagger}(\mathbf{r}) & H^{b}_{AB}
\end{pmatrix},
\end{equation}
where The continuum model of the AB stacked graphene of the top and bottom bilayer ($H^{t/b}_{AB}$) is given by, \begin{equation}\label{top_bernal}
    H^{t}_{AB}= \begin{pmatrix}
-iv_{F}\bm{\sigma}\cdot\bm{\nabla}_{\theta/2}+\frac{V}{2} & T_{AB}\\
T^{\dagger}_{AB} & -iv_{F}\bm{\sigma}\cdot\bm{\nabla}_{\theta/2}+\frac{V}{6}
\end{pmatrix}
\end{equation}
\begin{equation}\label{bottom_bernal}
H^{b}_{AB}=\begin{pmatrix}
-iv_{F}\bm{\sigma}\cdot\bm{\nabla}_{-\theta/2}-\frac{V}{6} & T_{AB}\\
T^{\dagger}_{AB} & -iv_{F}\bm{\sigma}\cdot\bm{\nabla}_{-\theta/2}-\frac{V}{2}
\end{pmatrix}
\end{equation}
An applied electric field perpendicular to the system results into a voltage difference between the bilayers ($V$ in equations (\ref{top_bernal}),(\ref{bottom_bernal})). The interlayer hopping within an AB stacked bilayer is captured in 
\begin{equation}
T_{AB}=\begin{pmatrix} 
0 & 0\\
\gamma & 0 
\end{pmatrix},
\end{equation}
where we only keep the direct hopping between the two layers with $\gamma$ being the strength of the hopping. We use the numerical value of $\gamma=0.361eV$ when we compute the band structure\cite{PhysRevB.89.035405,RevModPhys.81.109}.\\
The moire hopping between the bilayers is given by
\begin{equation}
\Tilde{U}(\mathbf{r})=\begin{pmatrix}
0 & 0\\
U(\mathbf{r}) & 0
\end{pmatrix},
\end{equation}
with $U(\mathbf{r})$ defined in equation (\ref{U(r)}).
Similar to TBG, The TDBG system also has a magic angle where the two bands near charge neutrality become (almost) flat. Upon applying an electric field perpendicular to the sample, these bands get gapped and can acquire a non zero valley Chern number with a rich phase diagram that depends on both the twist angle and the applied electric field\cite{PhysRevB.103.115201}. We will focus on the case where the valence and the conduction bands have Chern numbers $C=\pm2$. Time-reversal symmetry forces the bands in the opposite valley to have opposite Chern number so at charge neutrality the system is in the valley Chern insulator phase, as in TBG.

The lattice set up is the following; we start with two bilayers of AB stacked graphene directly on top of each other then we apply a relative twist $\theta$ between them. A semi-infinte ribbon is formed by taking the system to be infinite in one primitive superlattice direction (taken to be $\mathbf{A}_2$) and finite in the other. The resulting band structure of TDBG in a ribbon geometry shows qualitatively similar behavior to TBG where there lie edge states between the flat and the high energy bands. There are, however, no edge states that connect the two flat bands that would be associated with the valley Chern insulator phase. 
\begin{figure}[!t]
	\subfloat[\label{TDBG_CM}]{%
		\includegraphics[width=0.23\textwidth]{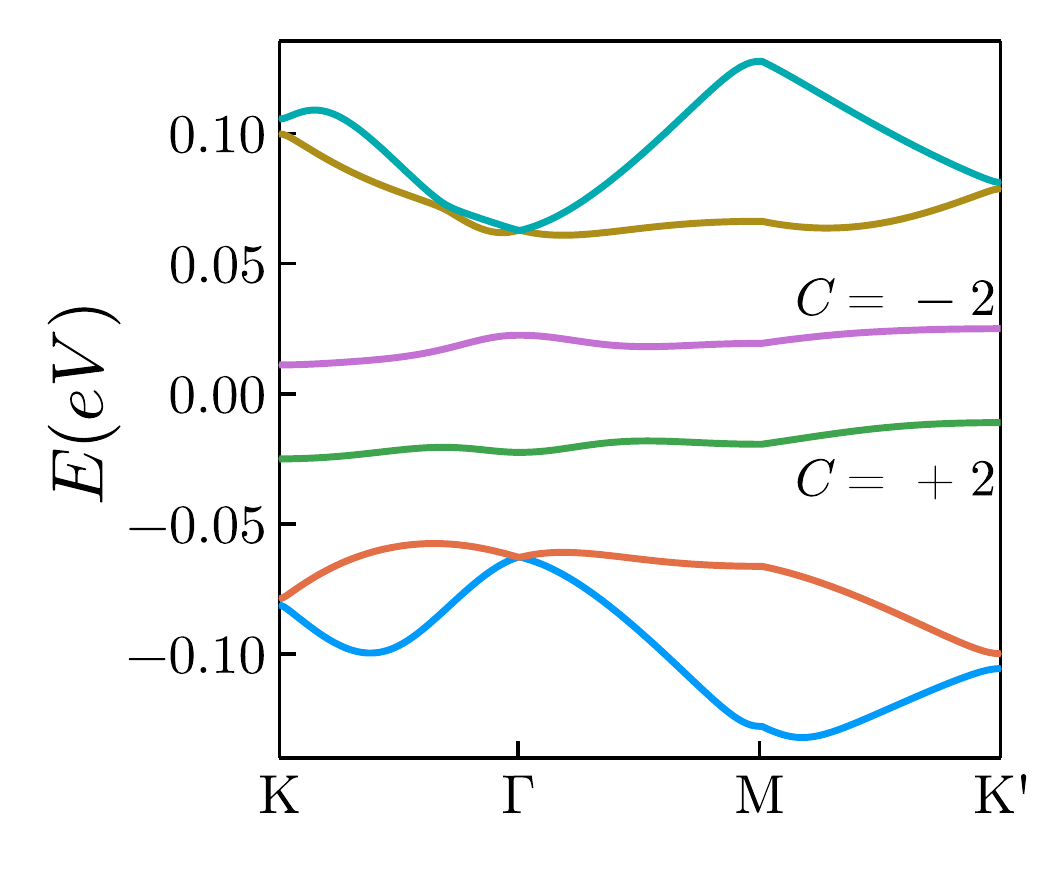}}
	\hspace{0.05em}
	\subfloat[\label{TDBG_lattice}]{%
		\includegraphics[width=0.23\textwidth]{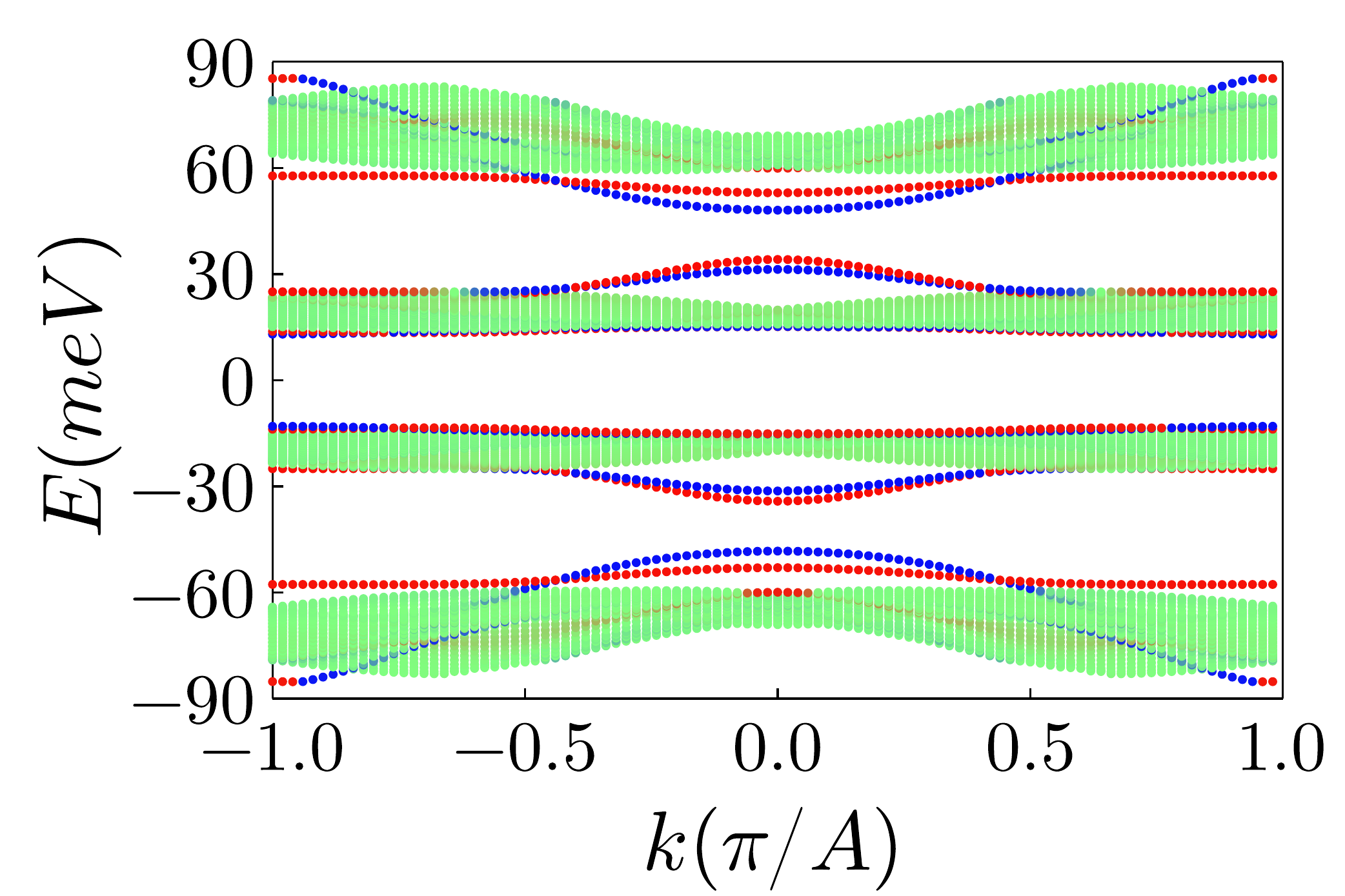}}\\
	\caption{\label{TDBG_ribbon}Continuum model and tight binding calculation in TDBG. (a) shows the continuum model band structure obtained at $\theta=1.3^\circ$ and $V=0.05 eV$. The two bands in the middle have nonzero Chern number which is twice that of TBG bands.(b) shows the band structure of a ribbon of width 20 moir\'e unit cells using the same parameters used in the continuum model and at a commensurate angle $\theta\approx1.3^\circ (m=26,n=25)$, $\kappa=0$ and $\xi=0.3$. Like TBG, we do not find gap crossing edge states near charge neutrality.}
\end{figure} 
\section{\label{sec5}Summary and Conclusions}
We have studied the continuum model for TBG aligned with HBN substrate and the TDBG system in a perpendicular electric field where single valley bands can acquire a nonzero Chern number in both systems. Due to the presence of time-reversal symmetry, opposite valleys have opposite Chern numbers making such systems candidates for realizing a valley Chern insulator with counter-propagating valley-polarized edge modes. We then studied the problem within a tight-binding model constructed on a lattice with the two layers rotated with respect to each other by a commensurate angle where Bloch theorem is applicable. We found that $C_3$ symmetry in the bulk does distinguish between the zero and the nonzero Chern number bands in the continuum as seen by a gap closing at the $K$ and $K'$ points in the moir\'e BZ that occurs at the transition point between the two phases. Moreover, the two touching bands exchange their $C_3$ eigenvalues at the critical point. There may be further interesting physics in this fine-tuned case, connected to the recently described "shift insulator"~\cite{shift_insulator} that we leave for future work.  We then solve the problem on a ribbon geometry where we can study the edge states of the system. Although edge states could be found between the flat bands and higher energy bands, edge states connecting the two flat bands at charge neutrality are absent. We attribute this to the mixing of the two valleys at the edge, and to the fact that generic edges will break the $C_3$ symmetry of the bulk. 

Our findings serve to intensify the puzzle posed by the experimental result of Ref.~\cite{Wang2022}. The experiment reports the observation of non-local transport, attributed to edge modes at charge neutrality in TDBG under the same conditions where we do not find edge modes. The physics missing from our model is that of electron-electron interactions. It is possible that the physical (sharp) edge is screened and modified by interactions in such a way as to reduce inter-valley scattering. An extension of our model, with the addition of Coulomb interactions even at the mean field level, can serve to investigate this possibility, which we leave for a future work.

\section*{ACKNOWLEDGMENTS}
The authors are grateful to A. Vishanwath and E. Khalaf for helpful discussions. This work was supported in part by NSF DMR-2026947 (AK, RKK). GM is grateful to the US-Israel BSF for partial support under grant no. 2016130. GM and RKK acknowledge the Aspen Center for Physics, NSF PHY-1607611 (GM, RKK) for its hospitality. The authors are grateful to the University of Kentucky Center for Computational Sciences and Information Technology Services Research Computing for their support and use of the Lipscomb Compute Cluster and associated research computing resources.
\appendix
\renewcommand{\theequation}{\Alph{section}.\arabic{equation}}
\renewcommand{\thefigure}{\Alph{section}.\arabic{figure}}
\setcounter{figure}{0}

\section{Comparison between the Continuum and the lattice model}
\label{app:lattcont}
\begin{figure}[ht!]
	\subfloat[\label{xi=0.1}]{%
		\includegraphics[scale=0.2]{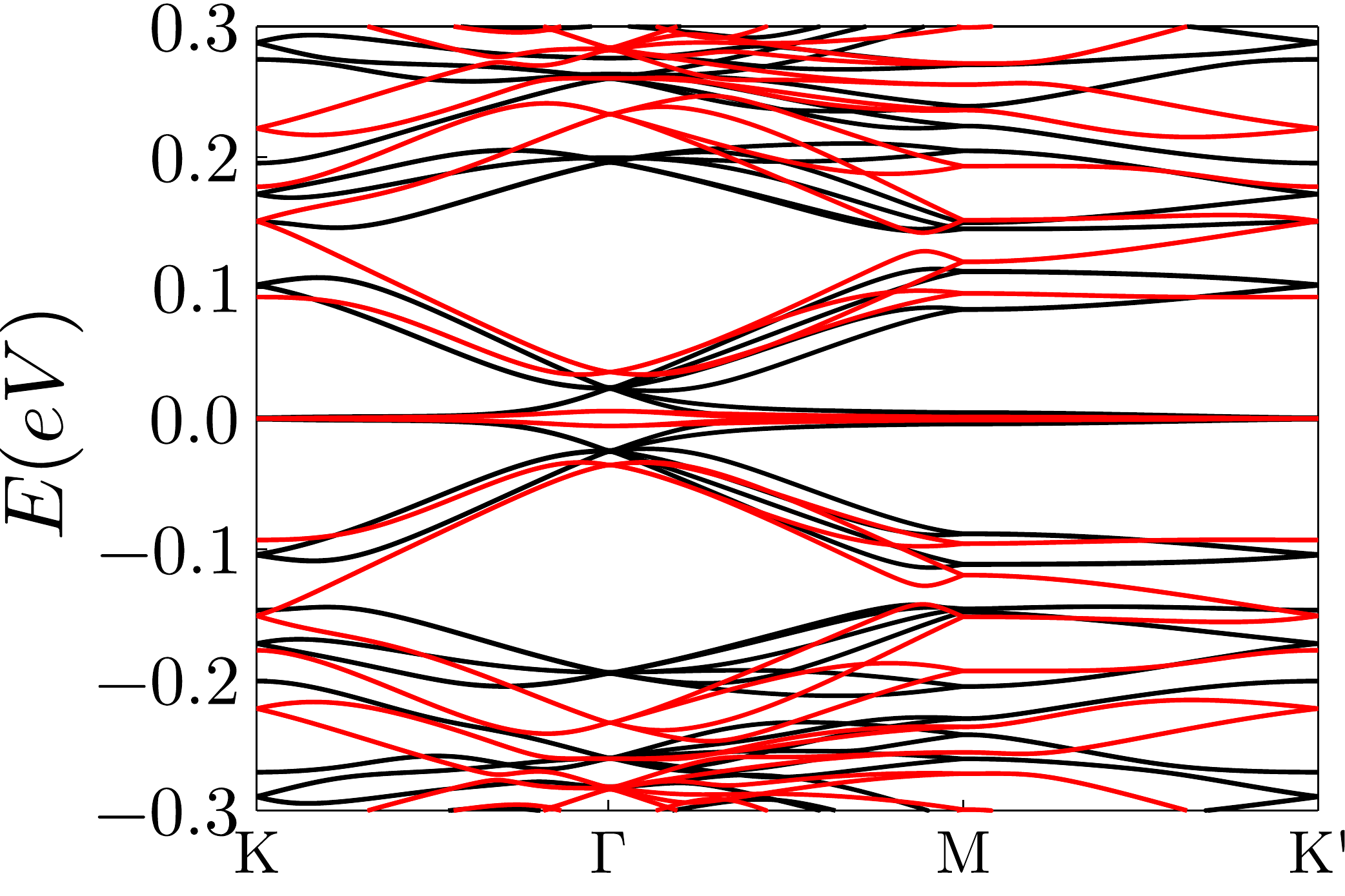}}
	\hspace{0.05em}
	\subfloat[\label{xi=0.3}]{%
		\includegraphics[scale=0.2]{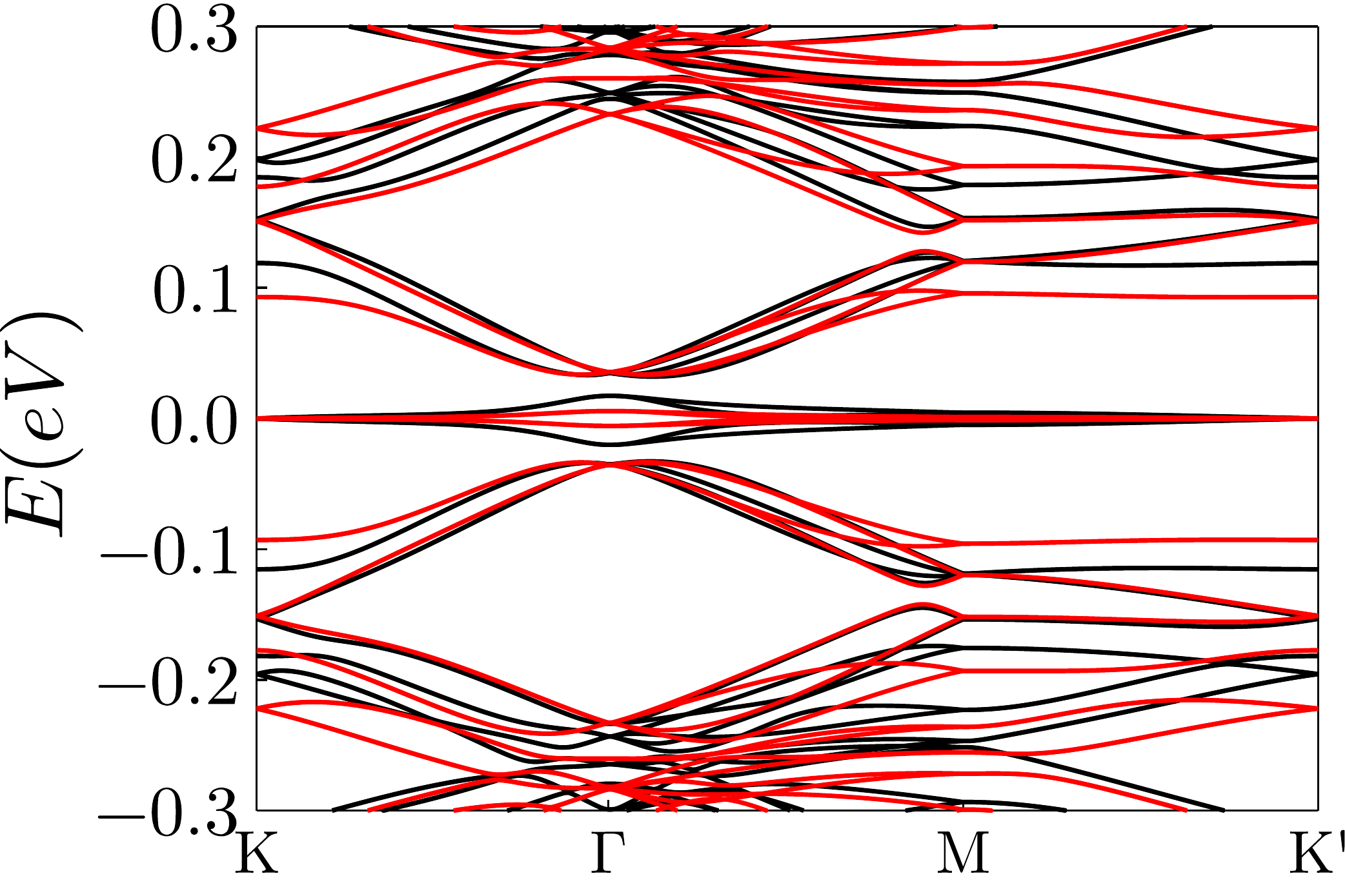}}
		\hspace{0.05em}
		\subfloat[\label{xi=0.6}]{%
		\includegraphics[scale=0.2]{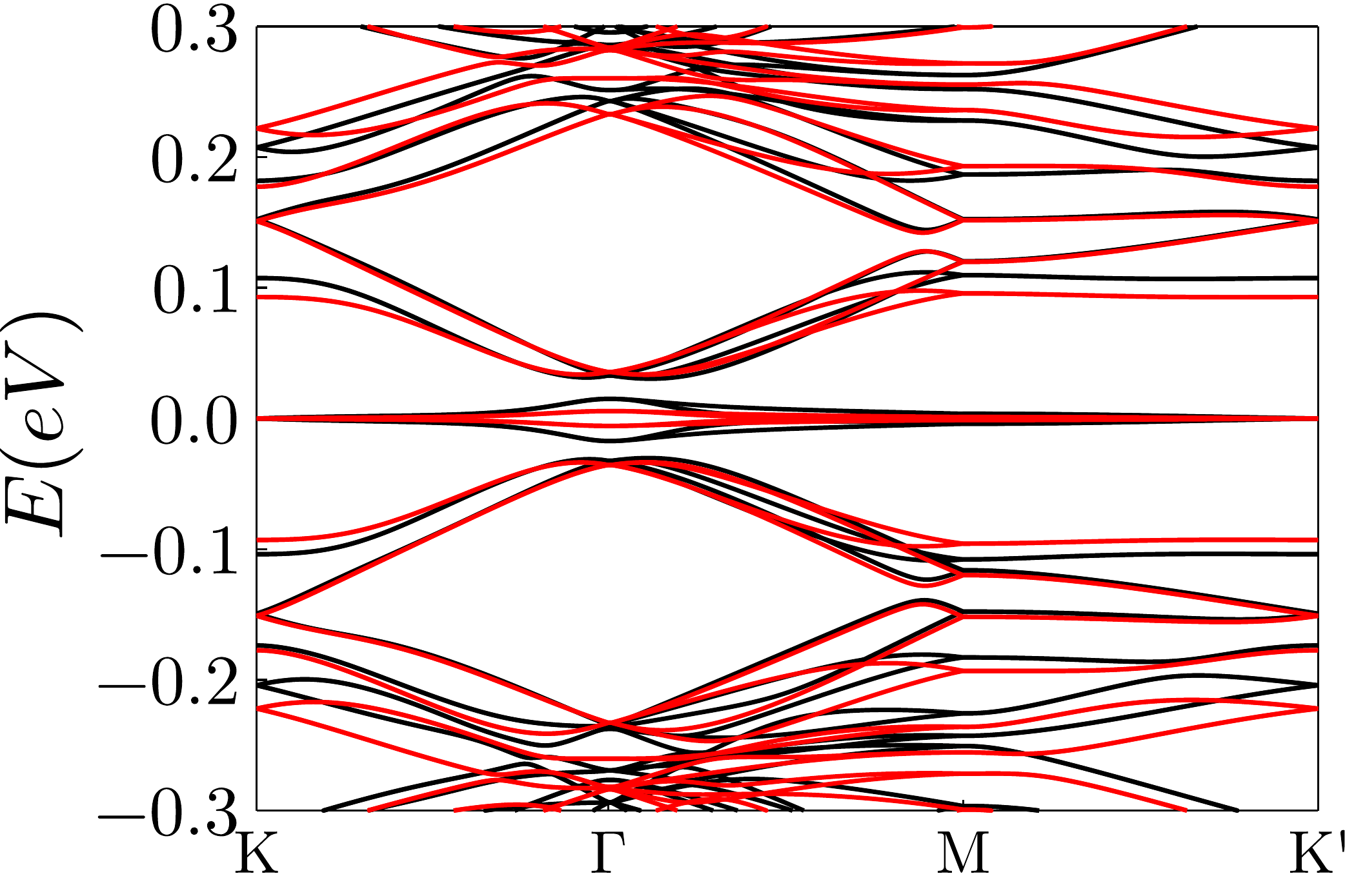}}\\
	\caption{\label{latticeTBG_vs_CM}Comparing the band structure obtained by the continuum and the lattice model in TBG. Both the twist angle and $\alpha$ are fixed at $\theta\approx1.08^\circ$($m=31,n=30$), and $\alpha=0.6051$. $\kappa=0.7$ was used. The tight binding model bands (black) are computed for different $\xi$ going from $\xi=0.1$ in (a), $\xi=0.3$ in (b) and $\xi=0.6$ in (c). The lattice model bands are converging to the continuum model (in red) as $\xi$ is increasing.}
\end{figure}

\begin{figure}[h!]
	\subfloat[\label{dxi=0.1}]{%
		\includegraphics[scale=0.2]{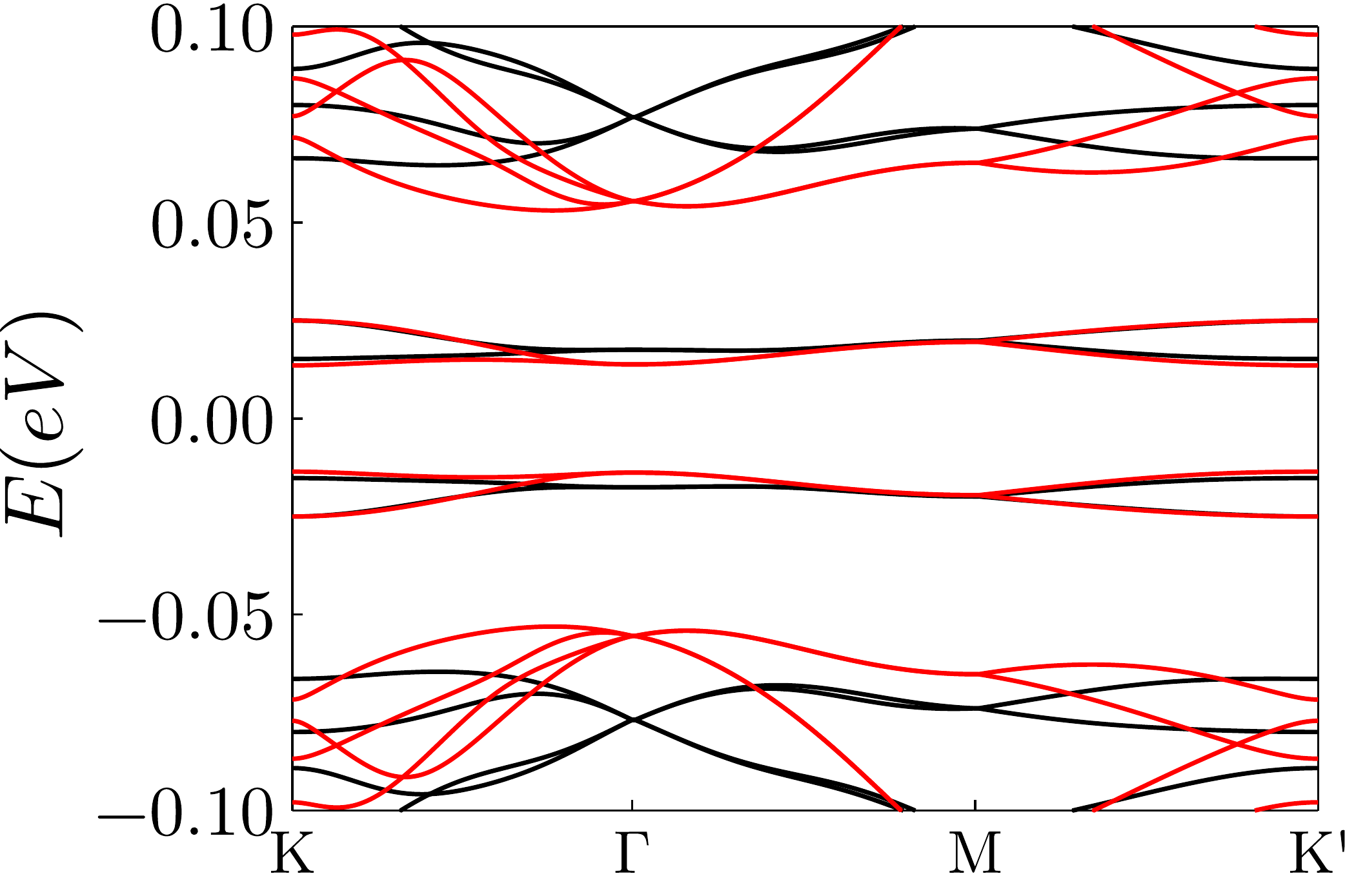}}
	\hspace{0.05em}
	\subfloat[\label{dxi=0.3}]{%
		\includegraphics[scale=0.2]{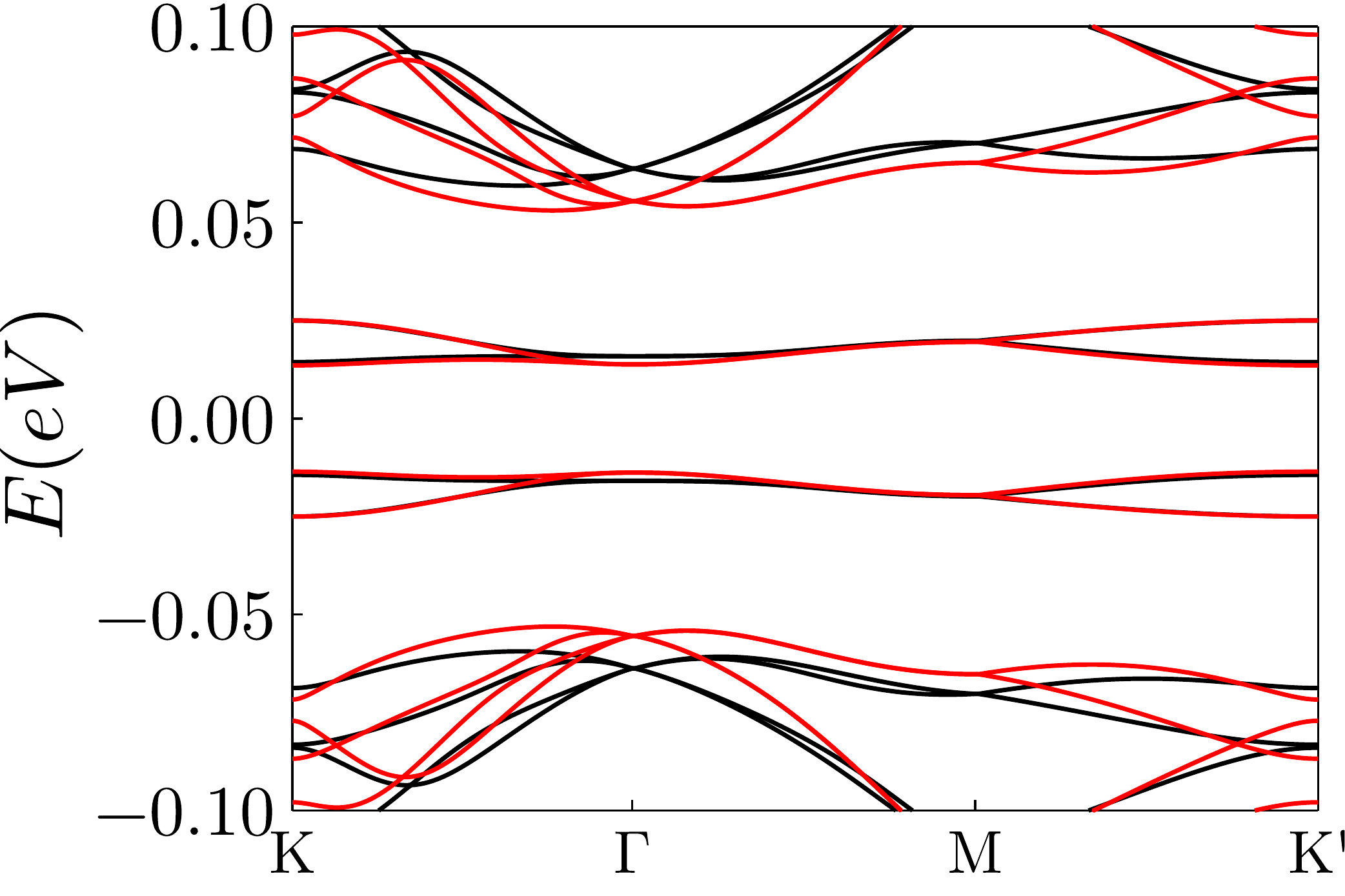}}
		\hspace{0.05em}
		\subfloat[\label{dxi=0.6}]{%
		\includegraphics[scale=0.2]{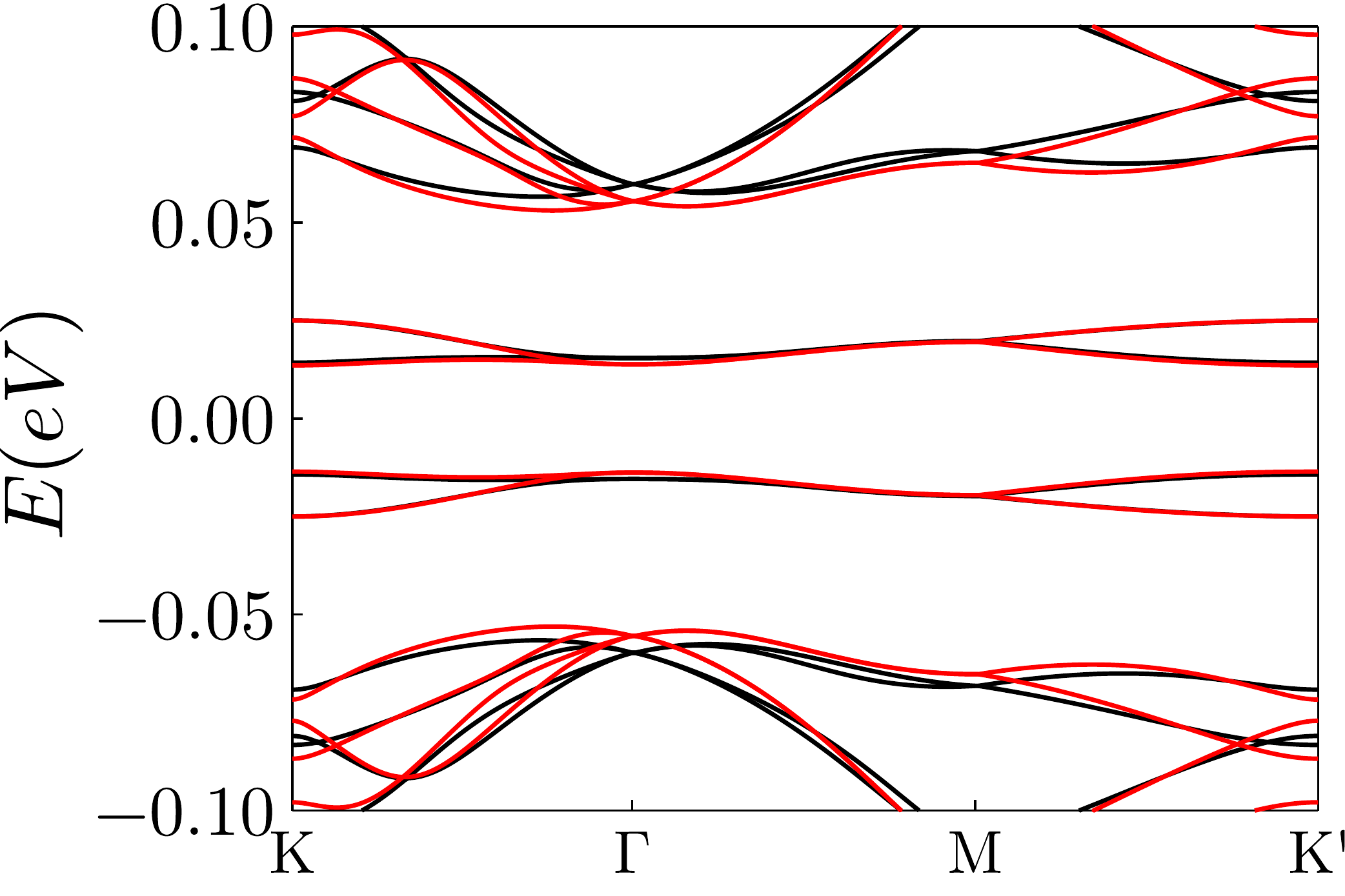}}\\
	\caption{\label{latticeTDBG_vs_CM}Comparing the band structure obtained by the continuum and the lattice model in TDBG. Both the twist angle and $\alpha$ are fixed at $\theta\approx1.3^\circ$ ($m=26,n=25$), and $\alpha=0.586$. The values of $\kappa=0.0$ and $V=50meV$ were used. The tight binding model bands (black) are computed for different $\xi$ going from $\xi=0.1$ in (a), $\xi=0.3$ in (b) and $\xi=0.6$ in (c). The lattice model bands are converging to the continuum model (in red) as $\xi$ is increasing.}
\end{figure}

In this appendix we show a comparison between the band structure obtained by our tight binding model at a commensurate angle and the continuum models of TBG and TDBG. This also serves as a way to verify our method. First we review the derivation of the inter layer hopping in the continuum model of TBG. One starts with Bloch waves in the two layers and then compute the Hamiltonian matrix element between the two Bloch states. The resulting matrix element takes the form: 
\begin{equation}\label{CM_matrixelem}
    T^{\alpha\beta}_{\bf{k}_t\bf{k}_b}=\sum_{\bf{G_t}\bf{G_b}}\Tilde{t}(\bf{k}_t+\bf{G}_t)e^{-i(\bf{k}_t+\bf{G}_t).(\bf{d}^{t}_{\alpha}-\bf{d}^{b}_{\beta})}\delta_{\bf{k}_{t}+\bf{G}_{t},\bf{k}_{b}+\bf{G}_{b}},
\end{equation}
where $\bf{k}_{t},\bf{k}_{b}$ are the momenta in the top and bottom layer, respectively, and $\alpha,\beta$ are labelling the graphene sublattices with $\bf{d}^{t}_{\alpha},\bf{d}^{b}_{\beta}$ denoting the position of the atoms within the unit cell. The sum is over the reciprocal lattice vectors ($\bf{G}_{t},\bf{G_{b}}$) of the top and the bottom layer. $\Tilde{t}(\bf{k}_t+\bf{G}_t)$ is the Fourier transform of the interlayer hopping function and the Kronecker delta ensures the conservation of momentum. 
One of the main assumptions of the continuum model is restricting the sum in (\ref{CM_matrixelem}) to the Dirac points in the graphene BZ. This rests on the fact that the hopping between the layers decays very rapidly with momentum\cite{PhysRevB.81.245412,doi:10.1073/pnas.1108174108}. Taking the form of the interlayer hopping given in the main text we have, 
\begin{equation}\label{hopping_Fourier}
    \Tilde{t}(q)=\int d^{2}r e^{-r/\xi}e^{-i\bf{q}.\bf{r}}=\frac{2\pi\xi^2}{(1+q^2\xi^2)^{3/2}}
\end{equation}
which shows $\Tilde{t}(q)$ is indeed a rapidly decaying function with momentum. The first order term that enters into the continuum model is $w_1=\Tilde{t}(k_D)/\Omega_{u.c}$. The next order hopping process beyond the the Dirac points has the momentum $q=2k_D$ with $k_D=4\pi/3$ is the momentum of the Dirac points in the graphene BZ (The lattice constant of graphene is set to 1). 
The ratio between the two amplitudes is given by,
\begin{equation}\label{amplitude_ratio}
\frac{\Tilde{t}(2k_D)}{\Tilde{t}(k_D)}=\left(\frac{1+\xi^2k_{D}^2}{1+4\xi^2k_{D}^2}\right)^{3/2}
\end{equation}
This shows that $\Tilde{t}(2k_D)$ is always less than $\Tilde{t}(k_D)$ and the ratio gets smaller as $\xi$ (the range of hopping in real space) gets bigger. There is a subtlety here, however, as one cannot keep increasing the range of the hopping uncontrollably as the continuum model is a low energy theory which assumes the linear dispersion of the bands. Having the range of the hopping too large will send the lattice theory beyond the low energy regime assumed in the continuum. With these restrictions in mind, we do a comparison between the band structure obtained via the lattice and the continuum model in both of TBG and TDBG. We show the bands of the lattice model converge to the continuum model bands as the range of the hopping increases (see Fig. \ref{latticeTBG_vs_CM} and Fig. \ref{latticeTDBG_vs_CM} ).
\section{Finite Size Scaling}
\setcounter{figure}{0}
In this appendix we support our numerical findings by examining how the ribbon band structure in TBG and TDBG evolves as the width of the ribbon increases. This way we gain confidence that what we find at finite width is going to survive in the thermodynamic limit. We obtain the band structure for ribbons with successive increase in their width. We start from 5 to 10 and then reach 20 moir\'e unit cells. We indeed see that the bands near charge neutrality remain virtually unchanged (see Fig. \ref{finite_size}) as the width of the ribbon is increased for both TBG and TDBG. The bands were obtained with $\kappa=0.0$ for simplicity. 
\begin{figure}[H]
	\subfloat[\label{TBG_w=5}]{%
		\includegraphics[scale=0.2]{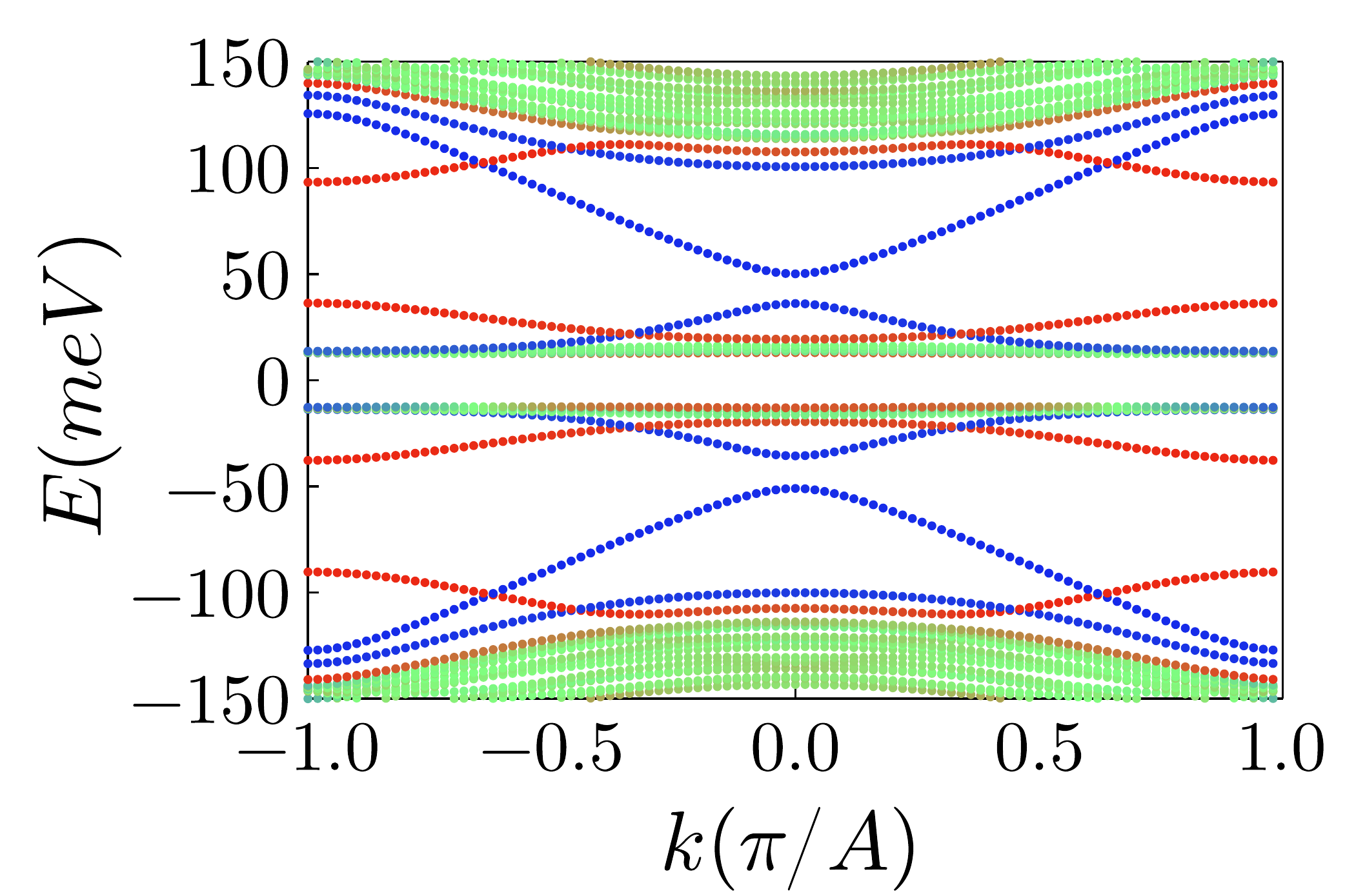}}
	\hspace{0.05em}
	\subfloat[\label{TDBG_w=5}]{%
		\includegraphics[scale=0.2]{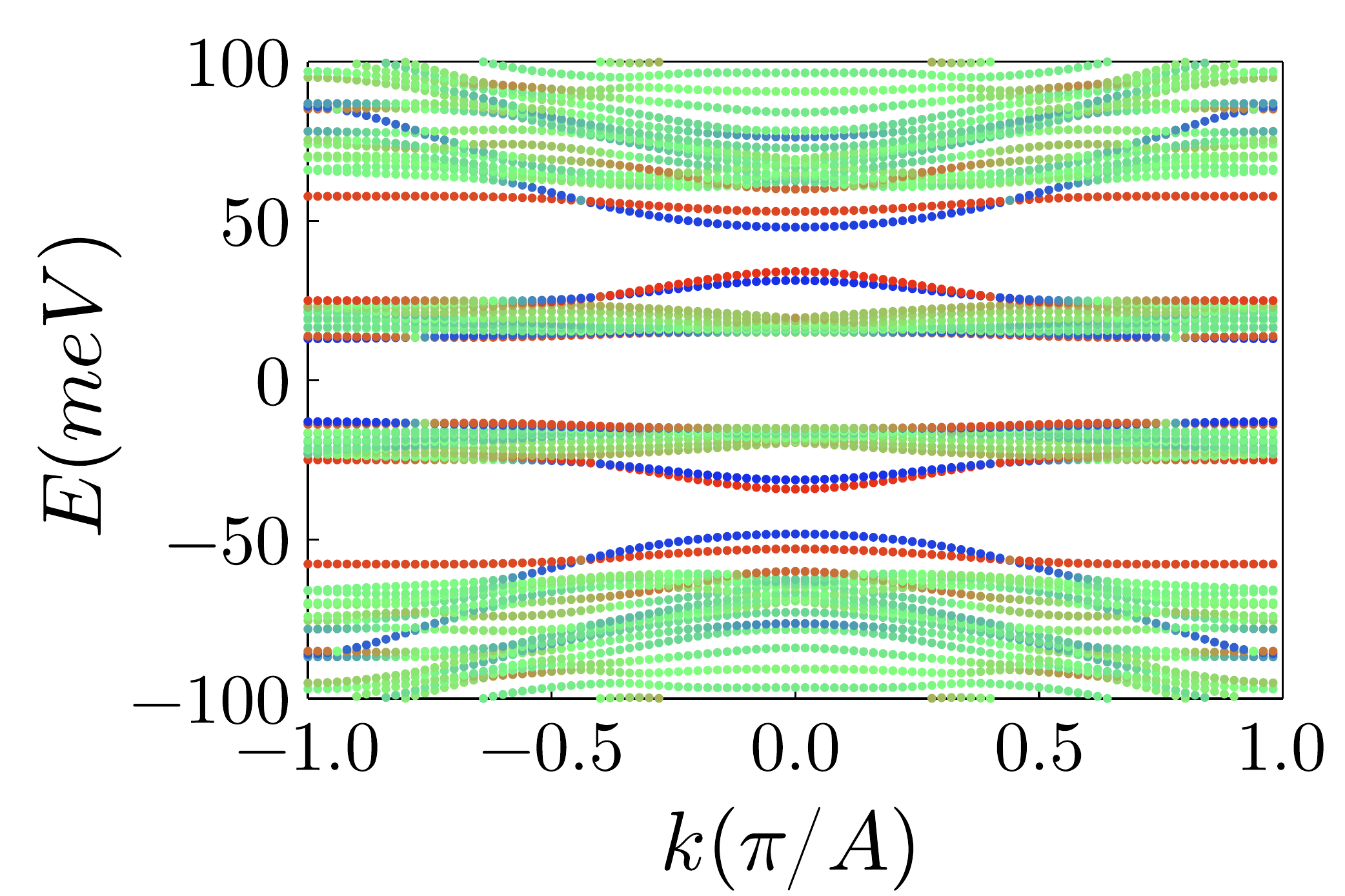}}
		\hspace{0.05em}
		\subfloat[\label{TBG_w=10}]{%
		\includegraphics[scale=0.2]{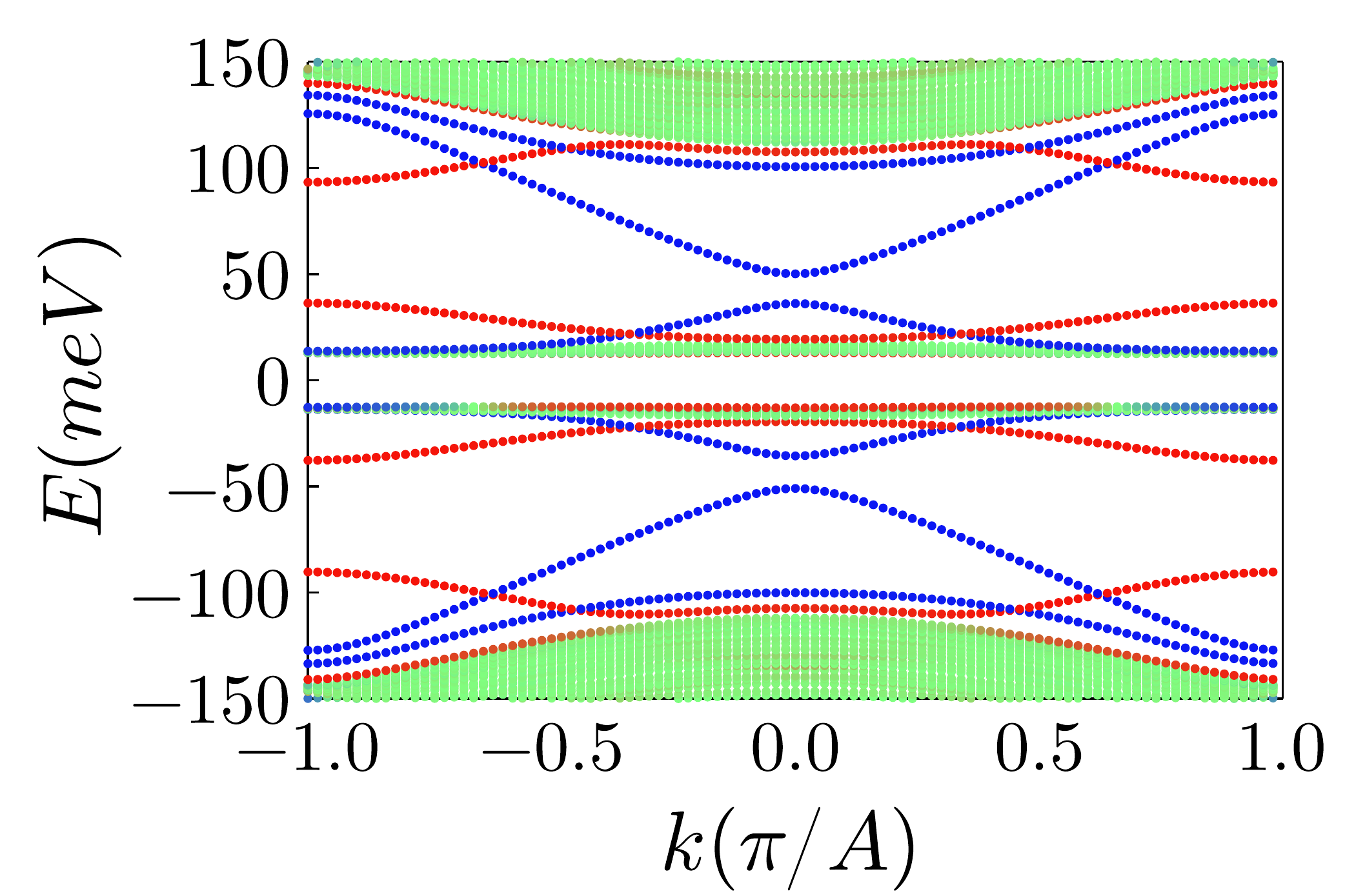}}
		\hspace{0.05em}
		\subfloat[\label{TDBG_w=10}]{%
		\includegraphics[scale=0.2]{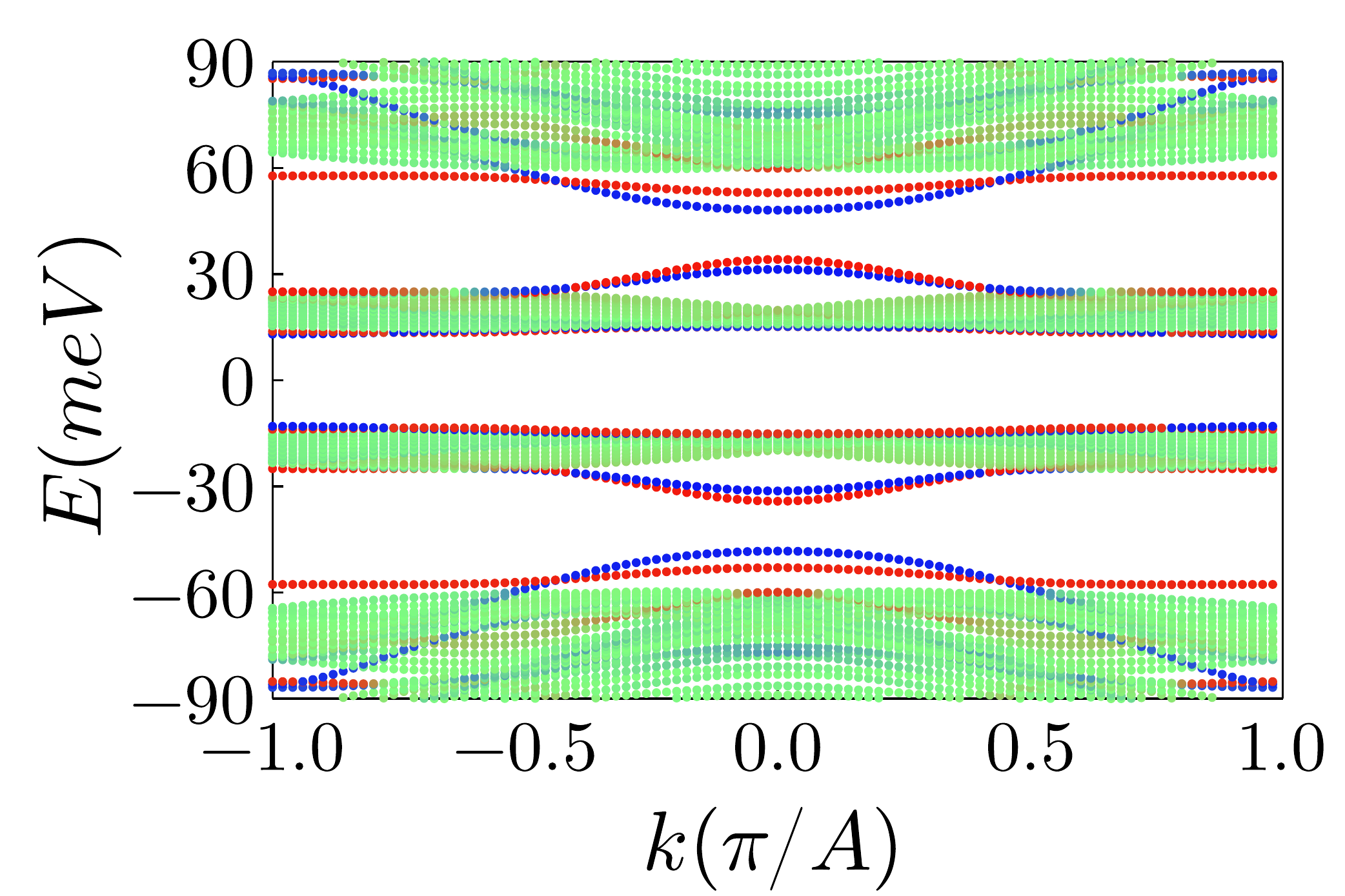}}
		\hspace{0.05em}
		\subfloat[\label{TBG_w=20}]{%
		\includegraphics[scale=0.2]{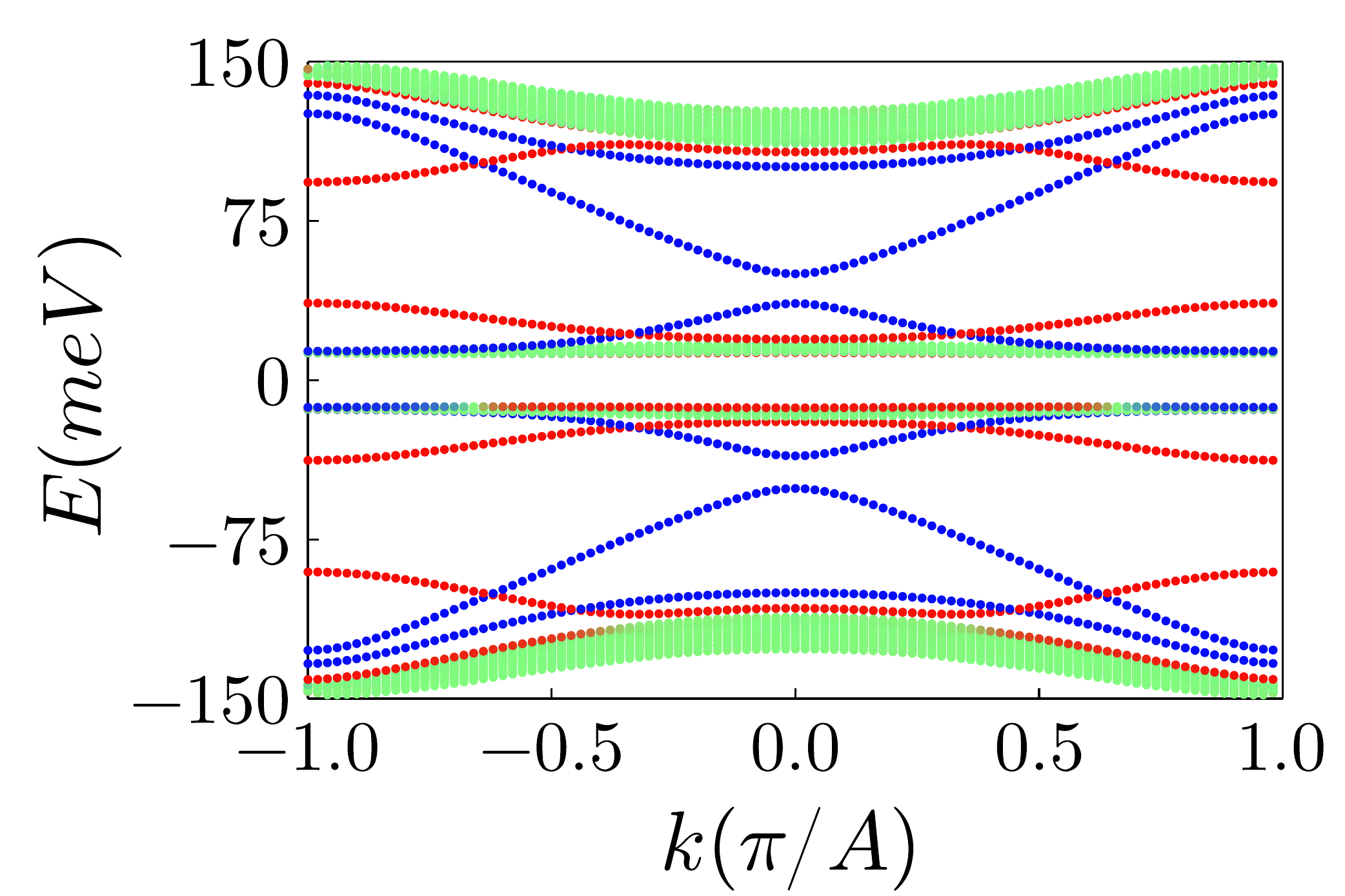}}
		\hspace{0.05em}
		\subfloat[\label{TDBG_w=20}]{%
		\includegraphics[scale=0.2]{dbands20.pdf}}\\
	\caption{\label{finite_size}Comparing the band structure as a function of the ribbon width. (a),(c),(e) shows the bands near charge neutrality in TBG obtained at a width of 5,10 and 20 moir\'e unit cells, respectively.(b),(d),(f) shows the bands near charge neutrality in TDBG obtained at a width of 5,10 and 20 moir\'e unit cells, respectively. In both systems we see that the bands do not change as a the width is increased implying a convergence to the thermodynamic limit.}
\end{figure} 
\bibliography{references}
\end{document}